\documentclass[11pt]{article} % use larger type; default would be 10pt
\pdfoutput=1 % if your are submitting a pdflatex (i.e. if you have
\usepackage{jheppub} % for details on the use of the package, please
\usepackage[utf8]{inputenc}
\usepackage[english]{babel}
\usepackage{amsmath,amssymb,amsbsy,amstext,amsthm,booktabs,simplewick,exscale,relsize,slashed,graphicx,amsfonts,upgreek, xcolor,subcaption}
\usepackage{multirow,color,dcolumn,bm,enumerate}
\usepackage{hyperref}
\usepackage{feynmp-auto,axodraw2,tikz}
\usepackage{ulem}
\usepackage{comment}

\usepackage[capitalise,noabbrev,nameinlink]{cleveref}

\newcommand{\half}{\frac 1 2}

\title{Constructing Operator Basis in Supersymmetry: A Hilbert Series Approach}
\author[a]{Antonio Delgado,}
\author[a]{Adam Martin,}
\author[a]{Runqing Wang}

\affiliation[a]{Department of Physics, University of Notre Dame,
  South Bend, IN, 46556 USA}

\emailAdd{adelgad2@nd.edu}
\emailAdd{amarti41@nd.edu}
\emailAdd{rwang7@nd.edu}

\abstract{
In this paper we introduce a Hilbert series approach to build the operator basis for a $N=1$ supersymmetry theory with chiral superfields. We give explicitly the form of the corrections that remove redundancies due to the equations of motion and integration by parts. In addition, we derive the maps between the correction spaces. This technique allows us to calculate the number of independent operators involving chiral and antichiral superfields to arbitrarily high mass dimension. Using this method, we give several illustrative examples.
}

\begin{document}
\maketitle

\setcounter{page}{2}

\section{Introduction}
\label{intro}

Effective field theory (EFT) is a framework which allows one to extend the Standard Model (SM)  to include new physics effects.  In EFT, higher dimensional operators encode our ignorance and also extends the theory up to a certain scale. There are many ways to form higher dimension operators by adding fields and derivatives but not all of these operators are independent. For example, within the Standard Model effective field theory it is difficult to determine the minimal operator basis above dimension six due to several redundancies in operator space \cite{Grzadkowski:2010es}. However, in the past few years, Hilbert series techniques have  provided an illuminating framework within which one is able to quickly count the number of independent operators at certain dimensions \cite{Lehman:2015via,Henning:2015daa,Lehman:2015coa,Henning:2017fpj,Anisha:2019nzx,Marinissen:2020jmb, Graf:2020yxt,Wang:2021wdq,Graf:2022rco,Yu2022,Yu2022a,Yu2021,Lu2021,Lu2022,Sun2022}. Several other methods, such as the on-shell Young Tableau construction~\cite{Henning:2019enq, Henning:2019mcv, Ma:2019gtx, Durieux:2019siw, Dong:2021yak,Li:2022tec,Fonseca:2019yya,Li:2020gnx}, also serve as complementary ways to give the number and explicit form of these operators. Using these tools, a complete list of independent dimension seven through nine operators can now be found in the literature \cite{Henning:2015alf,Murphy:2020rsh,Kobach:2017xkw,Gunawardana:2017zix,Liao:2020jmn,Li2021}.

Supersymmetry, as the largest spacetime symmetry compatible with an interacting theory, is still a possible candidate for physics BSM. Higher derivative operators in supersymmetry have been studied in several articles \cite{Antoniadis:2007xc,Forcella:2008tqe,Farakos:2013zsa,Dudas:2015vka, Nitta:2020gam} especially within the context of supersymmetry breaking. Therefore determining the operator basis in supersymmetry may be important, specially in order to avoid including operators that are related. However, there are difficulties that arise when applying the above mentioned EFT counting methods to a supersymmetric theory. The usual Hilbert series approach only works for Lorentz invariant as well as gauge invariant operator spaces. Once we include supersymmetry, it cannot directly give the correct counting, because the definition of integration by parts changes due to the additional fermionic derivatives in supersymmetric theories. 

In this paper we develop a method that allows one to build the operator basis of all dimensions in a $N=1$ supersymmetric theory with only chiral/antichiral superfields~\footnote{The inclusion of vector superfields will be postponed to a forthcoming publication.}. Hilbert series and related group theory techniques are  fundamental to tackle the problem in hand, and together with the definition of correction spaces (vector spaces of redundancies), provide a systematic way to remove all dependencies and get an operator basis. Using a recursive derivation, we are able to derive all corrections explicitly.

This paper is organized as follows. In Section \ref{review} we introduce the Hilbert series approach and how to apply it to form Lorentz invariants. Then we review $N=1$ supersymmetry in the language of superspace, and use Hilbert series tools to build the operator space we are interested in. Section \ref{relations and results} deals with the two kinds of redundancies, namely EOM and IBP relations. Specifically, in Section \ref{EOM} we show how to remove EOM. The method to remove IBP relations in non-supersymmetric theory is introduced in Section \ref{IBP in SMEFT}, leading to the definition of correction space, which allows one to identify corrections in a systematic way. We then generalize this idea and derive explicit corrections at each order in Section \ref{IBP in SUSY}, where three examples that involve chiral superfields and antichiral superfields are given at the end. Finally, in Section \ref{Conclusion} we give a brief summary and some possible future applications of this approach. Proofs and character formulae are given in the appendices\footnote{
Through out the paper we adopt the most-negative metric tensor in Minkowski space, i.e. $\eta^{\mu\nu}=\eta_{\mu\nu}=diag(+1, -1, -1, -1)$. Totally antisymmetric tensor in two dimensions $\epsilon_{AB}(A,B=1,2)$ are defined to be $\epsilon_{12}=\epsilon^{21}=1; \epsilon_{21}=\epsilon^{12}=-1$. In addition, a useful identity we will use is $\epsilon_{AB}\epsilon_{CD}+\epsilon_{AC}\epsilon_{DB}+\epsilon_{AD}\epsilon_{BC}=0$.}.

\section{Hilbert series and supersymmetry reviewed}\label{review}

\subsection{Hilbert series and plethystic exponential}\label{HS intro}

The Hilbert series~\cite{Pouliot:1998yv, Benvenuti:2006qr,Butti:2007jv,Feng:2007ur,Forcella:2007wk, Dolan:2007rq,Gray:2008yu,Hanany:2008sb,Benvenuti:2010pq,Chen:2011wn,Hanany:2012dm,Rodriguez-Gomez:2013dpa,Dey:2013fea,Hanany:2014hia,Hanany:2014dia} is a useful way to count the number of independent group invariants. In a field theory consisting of $N$ fields, the Hilbert series has the following form:
\begin{equation}
\mathcal{H}(D,\{\phi_a\})=\sum_{r_1,\cdot\cdot\cdot,r_N,k}c_{r_1,\cdot\cdot\cdot,r_N,k}\phi_1^{r_1}\cdot\cdot\cdot\phi_N^{r_N}D^k,
\end{equation}
where $c_{r_1,\cdot\cdot\cdot,r_N,k}$ is the number of independent invariants composed of $(r_1,\cdot\cdot\cdot,r_N)$ powers of $\phi_1\cdot\cdot\cdot\phi_N$ and $k$ derivatives. Technically, the $\phi_1\cdot\cdot\cdot\phi_N$ above are complex numbers to label the content of fields, i.e. spurions, and similarly $D$ is a complex number to label the partial derivative. To count the number of independent operators now becomes the same as calculating  $c_{r_1,\cdot\cdot\cdot,r_N,k}$.

One technique to calculate the Hilbert series is via the plethystic exponential (PE) ~\cite{Butti:2007jv, Gray:2008yu, Hanany:2008sb, Benvenuti:2010pq,Hanany:2014dia}. The plethystic exponential generates all (symmetric or anti-symmetric) products of its arguments. For our purposes, the arguments are spurions representing each field in the theory, multiplied by the character appropriate for that field's representation under the spacetime (Lorentz) and internal symmetries defining the theory. For a field $\phi_R$ transforming under the representation $R$ of a simple group $\mathcal G$, the plethystic exponential is defined as:
\begin{equation}
\label{eq:thePE}
PE[\phi_R ]=\exp\Big\{\sum_n\frac{1}{n}(\pm1)^{n+1}\phi_R^n\chi^n_{\mathcal G, R} \Big\},
\end{equation}
where we assign $+1$ for bosons (symmetric products) and $-1$ for fermions (antisymmetric products). The characters $\chi_{\mathcal G,R}$ can be expressed in terms of $\mathcal N$ unimodular complex variables, where $\mathcal N$ is the rank of $\mathcal G$. For example, $SU(2)$ is rank $1$, so all characters can be written in terms of a single complex variable $\alpha$. We will sometimes list the complex variables that make up the characters as arguments in the PE, e.g. $PE[\phi; \alpha]$ and refer to them as `group parameters'. Some explicit examples for $U(1)$ and $SU(2)$ characters are given in Appendix~\ref{app:characters}. A useful property of plethystic exponentials is:
\begin{equation}
PE[\phi_1]PE[\phi_2]=PE[\phi_1+\phi_2],
\end{equation}
which allows us to combine multiple fields into a single PE. 

Expanded, the PE is a sum over all polynomials of its arguments (fields, for us), with each term in the polynomial multiplied by some combination of characters. To project out the polynomials that are net gauge/Lorentz invariants, we use character orthonormality:
\begin{align}
\int d\mu_{\mathcal G}\, \chi^*_{\mathcal G, I}\, \chi_{\mathcal G,J} = \delta_{IJ},
\end{align}
where $d\mu_{\mathcal G}$ is the Haar measure of the group $\mathcal {G}$. The Haar measure can be treated as the group volume/measure defined on the group $\mathcal{G}$, and we give explictly its expression for common Lie groups in Appendix \ref{app:characters}. The above relation holds provided $\mathcal{G}$ is compact. Specifically, integrating the PE times $1$ -- the character of the trivial representation  -- over the Haar measure projects out $\mathcal G$ invariants and gives us the Hilbert series. For the example of $N$ fields $\phi$ in a theory defined by group $\mathcal G$ we get:
\begin{equation}\label{general hilbert series}
\mathcal{H}=\int d\mu_{\mathcal G}\, PE[\sum\limits_{i=1}^N\phi_i\chi_{R,i}^{}].
\end{equation}

If the theory has multiple symmetries, the argument of the PE is the product of the individual group characters, and the invariants are projected out by integrating over all Haar measures. When considering the symmetries of a theory, we include all gauge and internal global symmetries along with Lorentz symmetry. For the latter, we work with representations of $SU(2)_L \otimes SU(2)_R \cong SO(4)$ rather than $SO(3,1)$ since the former is compact and therefore its characters are orthonormal\footnote{We only care about counting the invariants and not about the dynamics.}. Since we have included the Lorentz group to the mix, we can add derivatives of fields to the PE, generating invariants of $\phi_R$, $\partial_\mu \phi_R$, etc. Naively, we can include fields with derivatives by adding them to the PE, meaning we treat e.g. $\partial_\mu \phi, \Box\phi$ as an independent field species and add them to the PE dressed with the appropriate characters. What this naive approach misses are redundancies among operators with derivaives from integration by parts and the equations of motion. These require special attention and will be addressed in detail shortly.

Forgetting about derivatives for the moment, let us calculate the (zero derivative) Hilbert series for $Q$ and $L$, two familiar left-handed fermions from the SM, as an example. Both transform as $(\frac 1 2,0)$ under $SU(2)_L \otimes SU(2)_R$, and $Q=\{3,2,\frac{1}{6}\}$  $L=\{1,2,-\frac{1}{2}\}$ under $\{SU(3)_c$, $SU(2)_W$, $U(1)_Y\}$. The argument of the plethystic exponential is given by:
\begin{equation}
\mathcal{I}(Q,L; x,y,u,z_1,z_2)=3Q(x+\frac{1}{x})(y+\frac{1}{y})(z_1+\frac{z_2}{z_1}+\frac{1}{z_2})u^{1/6}+3L(x+\frac{1}{x})(y+\frac{1}{y})u^{-1/2},
\end{equation}
where $x$ is the group parameter for $SU(2)_W$, $y$ is the group parameter for $SU(2)_L$, $u$ is the group parameter for $U(1)_Y$ and $z_1,z_2$ are group parameters for $SU(3)_c$. Notice that there is an additional factor of 3, which represents the 3 generations. Plugging $\mathcal{I}(Q,L; x,y,u,z_1,z_2)$ into the plethystic exponential integrating over all Haar measures, we get \cite{Lehman:2015via,Henning:2015alf}
\begin{equation}\label{smeft hilbert series}
\begin{split}
\mathcal{H}&=\int d\mu\, PE[\mathcal{I}(Q,L; x,y,u,z_1,z_2)]\\
&=1+57LQ^3+4818L^2Q^6+\cdot\cdot\cdot
\end{split}
\end{equation}
where $d\mu=d\mu_{SU(3)}(z_1,z_2)d\mu_{SU(2)}(y)d\mu_{SU(2)}(x)d\mu_{U(1)}(u)$\footnote{We could include the Haar measure for $SU(2)_R$ into $d\mu$ as well, with group parameter $w$. However, as none of the fields in the PE transform under $SU(2)_R$, the integral is trivial.}. We can therefore easily read out the number of operators from the expansion, i.e. 57 operators built from $LQ^3$ and 4818 operators built from $L^2Q^6$.

Having reviewed how to construct the (zero-derivative) Hilbert series for scalar and spinor fields using the plethystic approach, in the next section we will generalize this technique to supersymmetry, where the non-supersymmetric fields are replaced by superfields. Since our aim is to study the operator basis, we first need to know what kind of operators a supersymmetric Lagrangian can contain.

\subsection{$N=1$ supersymmetry}\label{N=1 SUSY}
In this section we introduce the basic knowledge of $N=1$ supersymmetry, restricting ourselves to (anti-)chiral superfields. To define such superfields, we need two superderivatives $D_\alpha,\ \overline{D}_{\dot{\alpha}}$, given by\footnote{The 4-dimensional sigma matrices are defined to be:
\begin{equation}
\sigma^{\mu}_{A\dot{B}}\equiv (I, \sigma_i);\ \overline{\sigma}^{\mu A\dot{B}}\equiv (I, -\sigma_i).
\end{equation}}:
\begin{subequations}\label{Ds}
\begin{align}
&D_{\alpha}=\frac{\partial}{\partial \theta^\alpha}-i\sigma^{\mu}_{\alpha\dot{\alpha}}\overline{\theta}^{\dot{\alpha}}\partial_\mu,\\
&\overline{D}_{\dot{\alpha}}=-\frac{\partial}{\partial\theta^{\dot{\alpha}}}+i\theta^\alpha \sigma^{\mu}_{\alpha\dot{\alpha}}\partial_\mu,
\end{align}
\end{subequations}
where $\theta_\alpha$ and $\overline{\theta}^{\dot{\alpha}}$ are two-dimensional Grassmann numbers, and $\partial_\mu$ is the usual partial derivative. From now on, we use $\partial_\alpha$ and $\partial_{\dot{\alpha}}$ to represent the two derivatives, i.e. $\partial_\alpha\equiv D_\alpha$ and $\partial_{\dot{\alpha}}\equiv \overline{D}_{\dot{\alpha}}.$ They satisfy the following anticommutation relation:
\begin{align}\label{anti relation}
\{\partial_\alpha,\partial_{\dot{\alpha}}\}=2i\sigma^{\mu}_{\alpha\dot{\alpha}}\partial_\mu.
\end{align}
The  (anti)commutation relations
\begin{equation}
\label{eq:useful}
[\partial_\beta,\{\partial_\alpha,\partial_{\dot{\alpha}}\}]=[\partial_{\dot{\beta}},\{\partial_\alpha,\partial_{\dot{\alpha}}\}]=0.
\end{equation}
will also prove useful in later sections.

Chiral superfields $\Phi$ and antichiral superfields $\Phi^\dagger$ are defined to satisfy the following constraints:
\begin{equation}
\partial_{\dot{\alpha}}\Phi=0,\ \partial_\alpha\Phi^\dagger=0.
\end{equation}
They can be expanded in terms of component fields:
\begin{align}
&\Phi=\phi(y)+\sqrt{2}\theta\psi(y)+\theta\theta F(y),\\
&\Phi^\dagger=\phi^*(y^\dagger)+\sqrt{2}\overline{\theta}\overline{\psi}(y^\dagger)+\overline{\theta}\overline{\theta} F^*(y^\dagger),
\end{align}
where $y^\mu=x^\mu-i\theta \sigma^\mu\overline{\theta}$ and $y^{\dagger\mu}=x^\mu+i\theta \sigma^\mu\overline{\theta}$ represent superspace coordinates, $\phi$ is a complex scalar field, $\psi$ a Weyl chiral fermion and $F$ an auxiliary field.

The supersymmetric action built from chiral and antichiral superfields is formed as \cite{Weinberg:2000cr}:
\begin{equation}
S=\int d^4x [(W(\Phi)+W^*(\Phi^\dagger))_\mathcal{F}+K(\Phi,\Phi^\dagger)_\mathcal{D}],
\end{equation}
where $W(\Phi)$ and $W^*(\Phi^\dagger)$ are holomorphic functions (superpotential)  of chiral and antichiral superfields respectively, and $K(\Phi,\Phi^\dagger)$ is a real scalar function of both $\Phi$ and $\Phi^{\dagger}$, called the K\"ahler potential. The subscripts $\mathcal{F},\mathcal{D}$ represent $\mathcal{F}$-term ($d^2\theta$ term) and $\mathcal{D}$-term ($d^2\theta d^2\overline{\theta}$ term) respectively. For example, the explicit renormalizable Lagrangian for a single chiral superfield $\Phi$ is given by:
\begin{equation}
\mathcal{L}=[(\frac{1}{2}m\Phi^2+\frac{1}{3}g\Phi^3)_\mathcal{F}+h.c.]+(\Phi\Phi^\dagger)_\mathcal{D},
\end{equation}
where the two functions are chosen to be $W(\Phi)=\frac{1}{2}m\Phi^2+\frac{1}{3}g\Phi^3$ and $K(\Phi,\Phi^\dagger)=\Phi\Phi^\dagger$.

To build the Hilbert series for $N=1$ supersymmetry with (anti)chiral superfields, we first need to know their characters. Since a chiral superfield contains both bosons and fermions, we choose the lowest component fields to represent $\Phi, \Phi^\dagger$ -- $\phi$ and $\phi^*$ respectively, which transform as scalar fields under the Lorentz group\footnote{Note that the full supermultiplet $S(x,\theta,\overline{\theta})$ can be built by acting supercharges on the lowest component field $A(x)$, i.e. $S(x,\theta,\overline{\theta})=e^{(\theta Q+\overline{\theta}\overline{Q})} A(x)$, where $Q$ and $\overline{Q}$ are group generators (supercharges) related to $N=1$ supersymmetry.}. If we were only interested in superfield invariants without superderivatives, Eq.~\eqref{general hilbert series} is sufficient. Operators with superderivatives are where all the complications arise and will be the main focus of the rest of this paper. 

%\footnote{This will be true for any odd number of superderivatives acting on $\Phi, \Phi^{\dag}$, while even numbers of superderivatives acting on $\Phi, \Phi^{\dag}$ are bosonic}
%$\partial_\beta\partial_{\dot{\alpha}}\partial_\alpha\Phi, \cdots$.

%To include fermions, we need superfields with odd numbers of superderivatives, e.g. $\partial_\alpha\Phi,\partial_\beta\partial_{\dot{\alpha}}\partial_\alpha\Phi, \cdots$. Their corresponding representations are also given by the lowest component fields. For example, the lowest component field of $\partial_\alpha\Phi$ is a left-handed fermion $\psi_\alpha$, which transforms as $(\frac{1}{2},0)$, as expected from the fact that $\partial_\alpha\sim(\frac{1}{2},0)$ and $\Phi$ transforms as $(0,0)$. Adding more superderivatives gives higher representations, e.g. $\partial_\beta\partial_\alpha\Phi$ transforms under $(0+1,0)$ representation. 

%  It is worth pointing out that to build the operator basis for (anti)chiral superfields with superderivatives is the same as to find independent $\mathcal{D}$-terms. 
%  In other words, if the $\mathcal{F}$-term contains any superderivatives $\partial_\alpha,\partial_{\dot{\alpha}}$, we are always able to transform it into a $\mathcal{D}$-term. 

Fields with one derivative, $\partial_\alpha\Phi$, $\partial_{\dot\alpha}\Phi^{\dag}$ have a lowest component that is fermionic. They carry Lorentz group representations $(\half,0)$, $(0,\half)$ respectively, and must be added to the fermionic portion of the PE, meaning they enter the sum in Eq.~\eqref{eq:thePE} with a minus sign. As in non-supersymmetric theories, we can add operators with more superderivatives to the PE, e.g. to generate even higher derivative operators. There are three differences with respect to the non-supersymmetric case in this aspect. First, certain terms are zero because of the chiral/antichiral nature of the $\Phi, \Phi^{\dag}$, e.g. $\partial_{\dot\alpha}\Phi$, and should not be added. Second, we have to keep track of the bosonic/fermionic nature of the higher derivative terms. This is straightforward, as all terms with an even number of superderivatives acting on $\Phi, \Phi^{\dag}$ are bosonic, while all terms with an odd number of superderivatives are fermionic. Finally, it may seem that we need to study higher derivative extensions of the superpotential and K\"ahler potential separately, as they have different holomorphy properties. However, as we will show, any $\mathcal{F}$-term (superpotential term) containing superderivatives $\partial_\alpha,\partial_{\dot{\alpha}}$ can be transformed into a $\mathcal{D}$-term (K\"ahler term). Therefore, to build our operator basis for (anti)chiral superfields with superderivatives, we only need to find the set of independent $\mathcal{D}$-terms.
  
 To prove the last statement, let $W=W(\Phi_i,\partial_{\dot{\alpha}}^2S_i)$, where $\Phi_i$ are chiral superfields that satisfy the chiral constraints $\partial_{\dot{\alpha}}\Phi_i=0$ and $S_i$ are general superfields. Since $\partial_{\dot{\alpha}}^3=0$ identically, $\partial_{\dot{\alpha}}^2S_i$ are chiral superfields. As a result, $W(\Phi_i,\partial_{\dot{\alpha}}^2S_i)$ constructed in this way is chiral and we can choose its $\mathcal{F}$-term to be part of the Lagrangian,
\begin{equation}
S\supset\int d^4xW(\Phi_i,\partial_{\dot{\alpha}}^2S_i)_\mathcal{F}+h.c.
\end{equation}
There are two kinds of terms that exist in $W(\Phi_i,\partial_{\dot{\alpha}}^2S_i)$, one is $W_1=W_1(\Phi_i)$ and another is $W_2=(\partial_{\dot{\alpha}}^2S_k) h(\Phi_i,\partial_{\dot{\alpha}}^2S_i)$, where $k=1,2,3,\cdot\cdot\cdot i$ labels one $S$ field. By definition, $W_1$ doesn't carry any derivatives and we can simply drop such terms. The other term, $W_2$, is the same as $W_2=\partial_{\dot{\alpha}}^2[S_kh(\Phi_i,\partial_{\dot{\alpha}}^2S_i)]$ since $\partial_{\dot{\alpha}}h(\Phi_i,\partial_{\dot{\alpha}}^2S_i)=0$ due to the chiral constraint. As a result, 
\begin{equation}
\begin{split}
&\int d^4xW_2(\Phi_i,\partial_{\dot{\alpha}}^2S_i)_\mathcal{F}\\
=&\int d^4x\{\partial_{\dot{\alpha}}^2[S_kh(\Phi_i,\partial_{\dot{\alpha}}^2S_i)]\}_\mathcal{F}\\
\sim&\int d^4x[S_kh(\Phi_i,\partial_{\dot{\alpha}}^2S_i)]_\mathcal{D},
\end{split}
\end{equation}
where we drop total derivatives in $x$-spacetime when we go from the second line to third line.  %From now on, we will only consider $\mathcal{D}$-term operators. 

We are then left to count the number of operators with an arbitrary number of superderivatives that can form a real function, the most general K\"ahler potential. To form an invariant, we need to form a Lorentz scalar, and therefore the number of superderivatives should be even to get fully contracted. In addition, due to the intrinsic existence of an R-symmetry -- a symmetry that transforms the $\theta$'s -- we have to put another constraint. We claim that an operator is R-invariant if it carries the same number of $\partial_\alpha$'s and $\partial_{\dot{\alpha}}$'s. This is easily proved by noticing that each $\partial_\alpha$ reduces one degree of $\theta$ (correspond to -1 to the R-charge), while each $\partial_{\dot{\alpha}}$ reduces one degree of $\overline{\theta}$ (correspond to +1 to the R-charge). If we assign the R-charge 0 to (anti)chiral superfields\footnote{For multiple flavours, we should assign R-charges $r_i$ to chiral superfields $\Phi_i$ of different flavours.}, the field $\partial_\alpha\Phi$ carries R-charge -1 while the field $\partial_{\dot{\alpha}}\Phi^\dagger$ carries R-charge 1. So the lowest dimensional R-invariant Lorentz scalar we can form out of $\partial_\alpha \Phi, \partial_{\dot\alpha}\Phi^{\dag}$ is $(\partial_\alpha\Phi)^2(\partial_{\dot{\alpha}}\Phi^\dagger)^2$. To incorporate R-symmetry into the Hilbert series, we need an additional $U(1)$ group as well as the related group parameter $z$. %and integrating over Haar measure for $U(1)(z)$ gives R-invariant terms.

Calculating the corresponding Hilbert series is straightforward by putting superfields together with their characters into the plethystic exponential and then integrating over the Haar measure. For example, suppose we want to form the Hilbert series with fields $\Phi, \Phi^\dagger, \partial_\alpha\Phi, \partial_{\dot{\alpha}}\Phi^\dagger$, whose representations are given by $(0,0;0),(0,0;0),(\frac{1}{2},0;-1),(0,\frac{1}{2};1)$ respectively; the last number represents the R-charge. Then the Hilbert series is formed as:
\begin{equation}\label{susy hs formula}
\int d\mu PE[\mathcal{I}(\Phi, \Phi^\dagger, \partial_\alpha\Phi, \partial_{\dot{\alpha}}\Phi^\dagger;\alpha,\beta,z)],
\end{equation}
where $\alpha,\beta,z$ represent the group parameters of $SO(4)$ and $U(1)_R$. The Haar measure in this case is $d\mu=d\mu_{SU(2)}(\alpha)d\mu_{SU(2)}(\beta)d\mu_{U(1)_R}(z)$, and the integrand in the PE is given by
\begin{equation}\label{susy hs example}
\begin{split}
&\mathcal{I}(\Phi, \Phi^\dagger, \partial_\alpha\Phi, \partial_{\dot{\alpha}}\Phi^\dagger;\alpha,\beta,z)\\
=&\Phi+\Phi^\dagger+(\partial_\alpha\Phi)(\alpha+\frac{1}{\alpha})z^{-1}+(\partial_{\dot{\alpha}}\Phi^\dagger)(\beta+\frac{1}{\beta})z.
\end{split}
\end{equation}
Plugging \eqref{susy hs example} into \eqref{susy hs formula} will generate all possible invariants constructed from these superfields.

Once we include fields with superderivatives in the PE, as in the example above, the Hilbert series generated by Eq.~\eqref{general hilbert series} (\eqref{susy hs formula} for the example just shown) is not the end of the story. Operators must be independent (represent different contributions to the action) in an operator basis, and we have not yet removed redundancies from the Hilbert series coming from IBP or EOM relations. The first redundancy comes from the fact that two operators differing by a total derivative gives the same action after integrating over the full space (assuming the boundary terms vanish). The second redundancy comes from field redefinition, after which the original operator can be replaced with another operator with fewer derivatives. We will discuss in detail how to eliminate these two relations in the next section.

%\textcolor{red}{
%One comment before moving on: Including terms with superderivatives in the K\"ahler potential may generate derivative terms for auxiliary terms which could introduce extra dynamics in the theory~\cite{Farakos:2013zsa}. It should be pointed out that the EOM of auxiliary fields is just $F=m \phi ^*$ and that identification will be used to eliminate any dependance of those fields. }

%Having new degrees of freedom could be problematic since, for the wrong sign kinetic term, they may destabilize the theory. However, any destabilizing effects will only be important at energies close to the cutoff of the EFT (the scale $\Lambda$ suppressing the higher dimensional terms). In the spirit of an EFT, we will not deal with any requirements that avoiding this (or other similar) issue place on the UV completion. Our interest is just to calculate the number of independent operators and not their effect.

\section{Removing EOM and IBP}\label{relations and results}

\subsection{EOM relations}\label{EOM}
To remove EOM relations from the operator space, we first look at the equation of motion of a free chiral superfield $\Phi$, given by:
\begin{equation}\label{superfield eom}
\partial_\alpha^2\Phi=m\Phi^\dagger,
\end{equation}
where $m$ is the mass of the superfield. One can verify this by expanding both sides in component forms and then compare the lowest components:
\begin{equation}
\partial_\alpha^2\phi=m\phi^*,
\end{equation}
where $\phi$ and $\phi^*$ are the lowest component fields of $\Phi$ and $\Phi^\dagger$. As expected, this reduces to the Klein-Gordon equation for a free complex scalar field. The equation of motion relations allow one to ``replace'' $\Box\phi \to \phi$, $\slashed{\partial}\psi \to \psi^\dag$, etc. within higher dimensional operators via field redefinitions \cite{Georgi:1991ch}. Extrapolating this logic to superfields, we can swap  factors of $\partial_\alpha^2\Phi$ for $\Phi^\dagger$, etc. within superfield operators. 

The next question is how to enforce this in forming the Hilbert series, i.e. automatically removing redundant operators by manipulating the plethystic exponential. We will proceed as in non-supersymmetric theories, following Ref.~\cite{Lehman:2015coa}. 

Specifically, using a scalar field theory as an example, we add $\partial_\mu \phi, \Box\phi, \partial^2_{\mu,\nu}\phi$ to the PE as separate terms. As the PE generates all possible polynomials of its arguments, this gets us polynomials of   $\partial_\mu \phi, \Box\phi, \partial^2_{\mu,\nu}\phi$ and so on. To account for the EOM, we simply exclude $\Box\phi$ from the PE, as any operator containing $\Box\phi$ can be transformed by field redefinition to an operator without the $\Box$, and thus already included in the operator counting. By the same logic, we drop $\partial^2\partial_\mu \phi, \Box^2\phi$ etc. from the PE.  Omitting these terms, we are left with only the symmetric derivatives at each order\footnote{Antisymmetric combinations of derivatives are either zero (in the case of ordinary derivatives), or a field strength $X_{\mu\nu}$ (if $\phi$ is charged under a gauge symmetry and the derivatives are covariant derivatives). In either case, the terms don't appear as the building blocks in PE.}, $ \partial_\mu \phi, \partial^2_{\{\mu,\nu\}}\phi$, etc., where $\{\cdot\cdot\cdot\}$ indicates traceless and symmetric pieces. In terms of characters, the PE argument for a scalar is
\begin{align}
\mathcal{I}(\phi,\ \partial_\mu\phi,\ \partial_{\{\mu,\nu\}}\phi,\ \cdot\cdot\cdot;\alpha, \beta,\mathcal{D}) &= \phi+D\,\phi\chi_{(\frac{1}{2},\frac{1}{2})}+D^2\,\phi \chi_{(1,1)}+\cdot\cdot\cdot) \nonumber \\ 
&=\phi(1+D\,\chi_{(\frac{1}{2},\frac{1}{2})}+D^2\,\chi_{(1,1)}+\cdot\cdot\cdot). 
\label{eq:derivscal}
\end{align}
Here, the characters refer to representations under $SU(2)_L \times SU(2)_R$ and $D$ is the spurion for the derivative, which we need to keep track of operator mass dimension; $\alpha,\beta$ are the group parameters related to $SU(2)_L$ and $SU(2)_R$.

For fermions, the process is the same -- we extend the PE to include derivatives, but omit $\slashed \partial\psi$ and its higher derivative counterparts. For example, for a left handed fermion $\psi_L$, the PE argument is
\begin{align}
\mathcal I(\psi_L, \partial_\mu \psi_L, \partial^2_{\{\mu,\nu\}}\psi_L, \cdots;\alpha, \beta,\mathcal{D}) = \psi_L(\chi_{(\frac 1 2,0)} + D\,\chi_{(1, \frac 1 2)} + D^2\,\chi_{(\frac 3 2,1)} + \cdots ).
\label{eq:derivferm}
\end{align}
While we will not consider field strengths in this paper, one can account for their EOM in a similar fashion~\cite{Lehman:2015via}.

The infinite series of higher derivatives in Eq.~\eqref{eq:derivscal}, \eqref{eq:derivferm} can be summed. The results are, respectively, the characters for the scalar and fermion (here, $(\frac 1 2, 0)$ type)\footnote{Technically, and importantly for the approach in Ref.~\cite{Henning:2015daa,Henning:2017fpj}, the characters one gets by summing the infinite series of derivatives are short representations of the conformal group.} {\it conformal group} representations~\cite{Dolan:2008vc,Dolan:2005wy,Dolan:2002zh}. We'll denote the conformal representations as $\bar{\chi}_{(0,0)}$ and $\bar{\chi}_{(\frac 1 2,0)}$, so that Eq.~\eqref{eq:derivscal}, \eqref{eq:derivferm} can be expressed concisely as
\begin{align}
\mathcal I(\phi;\alpha, \beta, D) = \phi\,\bar{\chi}_{(0,0)},\quad \mathcal I(\psi_L;\alpha, \beta, D) = \psi_L\,\bar{\chi}_{(\frac 1 2,0)}.
\end{align}
Taking $\alpha$ and $\beta$ to be the group parameters for $SU(2)_L$ and $SU(2)_R$, the conformal characters are explicitly given by 
\begin{align}
\bar{\chi}_{(0,0)} &= P(\alpha, \beta,D)(1-D^2) \nonumber \\
\bar{\chi}_{(\frac 1 2,0)} &= P(\alpha,\beta,D )((\alpha+\frac{1}{\alpha})-D(\beta+\frac{1}{\beta})) \nonumber \\
\bar{\chi}_{(0,\frac{1}{2})} &=P(\alpha,\beta,D)((\beta+\frac{1}{\beta})-D(\alpha+\frac{1}{\alpha})), 
\label{eq:confchar}
\end{align}
where
\begin{align}
P(\alpha, \beta,D) = \Big( (1-D\alpha\beta)(1-\frac{D}{\alpha\beta})(1-\frac{D\alpha}{\beta})(1-\frac{D\beta}{\alpha}) \Big)^{-1}.
\end{align}
Notice that the conformal characters contain spurion $D$ along with the Lorentz group characters. The connection of the conformal group is not coincidental and has been used in Ref.~\cite{Henning:2017fpj} to analyze the Hilbert series for non-supersymmetric theories. 

The non-supersymmetric approach to EOM redundancy can be imported almost as is to the (chiral field) supersymmetric case. The complications are that i.) in supersymmetry we always have bosonic and fermionic fields, and ii.) there are two types of derivative. Both are easy to accommodate. For the two derivative types, we use $P$ as the spurion for $\partial_\alpha$ and $Q$ for $\partial_{\dot{\alpha}}$ -- the connection between $\partial_\alpha, \partial_{\dot{\alpha}}$ and $\partial_\mu$ from Eq.~\eqref{anti relation} implies $P Q \sim D$. For $\Phi$ and $\Phi^{\dag}$, we add derivatives following Eq.~\eqref{eq:derivscal}, simply substituting $PQ$ for $D$\footnote{The order of the two spurions P,Q doesn't matter since they are not real quantum operators.}. Ignoring $R$ symmetry for the moment,
\begin{align}
\mathcal I (\Phi, \Phi^{\dag};\alpha, \beta, P,Q) &= (\Phi+\Phi PQ\chi_{\frac{1}{2},\frac{1}{2}}+\Phi P^2Q^2\chi_{1,1}+\cdot\cdot\cdot) + \text{same for}\, \Phi^{\dag} \nonumber \\
&\equiv \Phi\bar{\chi}_{(0,0)} + \Phi^{\dag}\bar{\chi}_{(0,0)}
\end{align}
where it is understood that the arguments of $\bar{\chi}$ for the supersymmetric case are $P,Q$ and the group parameters $\alpha$ and $\beta$. For the fermionic fields $\partial_\alpha\Phi$, $\partial_{\dot{\alpha}}\Phi^{\dag}$, we follow Eq.~\eqref{eq:derivferm},
\begin{align}
\mathcal I(\partial_\alpha\Phi, \partial_{\dot{\alpha}}\Phi^{\dag};\alpha, \beta, P,Q) & = \partial_\alpha \Phi\, P\, \chi_{(\frac 1 2,0)} + \partial_\alpha \Phi\, P^2Q\,\chi_{(1,\frac 1 2)} + \partial_\alpha \Phi\, P^3Q^2\,\chi_{(\frac 3 2,1)}  \cdots   \nonumber\\
&\quad\quad + \partial_{\dot{\alpha}} \Phi^{\dag}\, Q\, \chi_{(0,\frac 1 2)} + \partial_{\dot{\alpha}} \Phi^{\dag}\, PQ^2\,\chi_{(\frac 1 2,1)} + \partial_{\dot{\alpha}} \Phi^{\dag}\, P^2Q^3\,\chi_{(1,\frac 3 2)} \cdots  \nonumber \\
&\equiv \partial_\alpha \Phi\, P\, \bar{\chi}_{(\frac 1 2,0)} + \partial_{\dot{\alpha}}\Phi^{\dag}\, Q\, \bar{\chi}_{(0,\frac 1 2)}
\label{eq:PEfermEOM}
\end{align}
The extra factors of $P$ and $Q$ in the last line of Eq.~\eqref{eq:PEfermEOM} account for the fact that the fermionic fields $\partial_\alpha \Phi, \partial_{\dot{\alpha}}\Phi^{\dag}$ already contain one superderivative.  As there are two types of derivative, it may not be obvious that only symmetric derivative combinations should be included in the fermionic PE. To see why, consider the example $\partial_\alpha{\partial }_{\dot{\alpha}}\partial ^\alpha\Phi$, which is the same as  $2i\sigma^{\mu}_{\alpha\dot{\alpha}}\partial_\mu\partial^\alpha\Phi$ when we anticommute the first two superderivatives and remove the piece that contains $\partial^\alpha\partial_\alpha\Phi$. However, expanding $\partial_{\alpha\dot{\alpha}}\partial^\alpha\Phi$ in component form we get $i{\sigma}^\mu_{\alpha\dot{\alpha}}\partial_\mu\psi^\alpha=m\psi^{\dagger}_{\dot{\alpha}}$ --  exactly the Dirac equation.

Putting the pieces together for a single chiral superfield (and its hermitian conjugate) and reinstating the $R$ symmetry with $R[\Phi] = r$, the full PE is
\begin{align}
& PE[\mathcal I (\Phi, \Phi^{\dag};\alpha, \beta, P,Q)]PE[\mathcal I(\partial_\alpha\Phi, \partial_{\dot{\alpha}}\Phi^{\dag};\alpha, \beta, P,Q)] = 
\nonumber \\
&\quad\quad  PE[\Phi\, z^r\, \bar{\chi}_{(0,0)} + \Phi^{\dag}\, z^{-r}\, \bar{\chi}_{(0,0)}]PE[\partial_\alpha \Phi\, z^{r-1}\,P\, \bar{\chi}_{(\frac 1 2,0)} + \partial_{\dot{\alpha}}\Phi^{\dag}\, z^{1-r}\, Q\, \bar{\chi}_{(0,\frac 1 2)}]
\end{align}
where $z$ is the $U(1)_R$ group parameter. Replacing $\Phi \to \sum_i \Phi_i$ , the PE can be extended to more $R[\Phi_i] = r_i$ chiral superfields. To account for fields with different $R$ charges, the full PE is the product over the individual $R$-charge sectors.

Integrating the PE over the Haar measure for the Lorentz group, $U(1)_R$, and any additional gauge/internal symmetry groups, the resulting Hilbert series includes derivatives and accounts for EOM redundancies. IBP redundancies are more subtle, and will be explored in detail in the next section. 

Before moving on, it is worth noting that while the Hilbert series contains all invariants, we are often only interested in invariants for a specific mass dimension. To address this, we can weight each spurion in the PE (both fields and derivative spurions) by their canonical mass dimension, e.g. $\Phi \to \epsilon \Phi, \partial_\alpha \Phi \to \epsilon^{3/2}\partial_\alpha\Phi$, etc. then expand the PE to the desired $\epsilon$ order before integrating over the Haar measure. This not only allows us to organize the invariants by mass dimension, but it simplifies the contour integration over the group parameters greatly as the only residues after expanding in $\epsilon$ are at the origin.

Finally, as most phenomenological applications of supersymmetry involve renormalizable operators only, it is worth spending a little more time on the meaning of higher dimensional superfield operators. Theories with higher dimensional superfield operators come about from integrating out fields fully supersymmetrically, and can be arrived at by performing the path integral over heavy degrees of freedom~\cite{Affleck:1984xz, Intriligator:1995au, Brizi:2009nn}. If all operators (including superderivatives) are maintained at a certain mass dimension in the expansion, the theory in terms of superfield is guaranteed to be supersymmetric (up to even higher dimensional effects). While convenient, superfields contain auxiliary fields, which seem confusing at first when present in higher dimensional operators. However, within the basis selected by the Hilbert series -- where as many derivatives as possible are removed via EOM -- the auxiliary fields do not become dynamical. As such, if one wants to convert between a higher dimensional superfield operators into its components, we can remove auxiliary fields (again, up to even higher dimensional effects) by the component form of the EOM, $F =  m\, \phi^*$ (if a mass term is allowed by the $R$ charges). 

\subsection{IBP relations in non-supersymmetric theories}\label{IBP in SMEFT}
With the EOM relations taken care of, in this section we will study the IBP redundancies. We begin by reviewing how IBP redundancies are handled in Hilbert series for non-supersymmetric field theories. As we will show, the structure of the IBP corrections in the non-supersymmetric case will guide us towards a generalization that works for supersymmetry.

For non-supersymmetric field theories, IBP redundancies can be accounted for by adding a factor to the Haar measure integrand, Eq.~\eqref{general hilbert series}~\cite{Henning:2015alf}, 
\begin{align}
\mathcal{H}=\int d\mu \frac 1 {P}\, PE[\sum_i \phi_i\chi_{R,i}],
\end{align}
where $P$ is the same function of the derivative spurions and Lorentz group parameters $\alpha$ and $\beta$ that we saw in the conformal characters (Eq.~\eqref{eq:confchar}), 
\begin{equation}\label{1/p}
P(D,\alpha,\beta)=\frac{1}{(1-D\alpha\beta)(1-\frac{D}{\alpha\beta})(1-\frac{D\alpha}{\beta})(1-\frac{D\beta}{\alpha})}.
\end{equation}
To understand the how $\frac{1}{P(D,\alpha,\beta)}$ incorporates IBP relations, let's expand it. Grouped by powers of $D$, $1/P$ is the sum of five terms:
\begin{equation} \label{expansion of P}
\begin{split}
\frac{1}{P(D,\alpha,\beta)}&=(1-D\alpha\beta)(1-\frac{D}{\alpha\beta})(1-\frac{D\alpha}{\beta})(1-\frac{D\beta}{\alpha})\\
&=1-D(\alpha+\frac{1}{\alpha})(\beta+\frac{1}{\beta})+D^2[(1+\alpha^2+\frac{1}{\alpha^2})+(1+\beta^2+\frac{1}{\beta^2})]\\
&-D^3(\alpha+\frac{1}{\alpha})(\beta+\frac{1}{\beta})+D^4.
\end{split}
\end{equation}
Plugged into the Haar measure integral, character orthonormality will project out different terms for each power of $D$. The first term, $\mathcal O(D^0)$ , is the same as what we had without the factor of $1/P$ -- it is the number of invariant operators and therefore sits in the $(0,0)$ Lorentz representation. Going forward, we'll refer to this set of operators as $\{ X \}$. The second term comes with a minus sign and accompanies the character for the $(\frac 1 2, \frac 1 2)$ Lorentz representation, therefore when we perform the Haar integral we'll project out all operators that are invariant under any internal/gauge symmetries but are Lorentz four-vectors -- the operator set $\{ X ^\mu\}$. By same logic, the $\mathcal O(D^2)$ term projects out all $(0,1) + (1,0)$ Lorentz representations, etc.

What does this have to do with IBP? Imagine we are looking at a theory of a single real scalar and care about counting invariant operators of the form $\mathcal O(\partial^m \phi^n)$.  IBP relations manifest here in the ways we can shuffle how the derivatives are sprinkled among the fields, with two operators being equivalent if they differ only by a total derivative. The operators projected out by the $\mathcal O(D)$ term in Eq.~\eqref{eq:ibpsimple} have the form $\mathcal O(\partial^{m-1} \phi^n)$. If we apply one final derivative to any of the $\mathcal O(\partial^{m-1} \phi^n)$ operators, we have to get zero since it's a total derivative. At the same time, $\partial_\mu [ \mathcal O(\partial^{m-1} \phi^n) ]$ must yield combination of $\mathcal O(\partial^m \phi^n)$ operators. So, for every $O(\partial^{m-1} \phi^n)$ operator, we find some linear combination of $O(\partial^m \phi^n)$ operators that equals zero; and for each linear combination, we can solve for one of the $\mathcal O(\partial^m \phi^n)$ operators in terms of the others, meaning it is not independent. For a more general (non-supersymmetric) theory, we can express the IBP relation as 
\begin{equation}
\partial_\mu X^\mu=\sum_i a_i X_i.
\label{eq:ibpsimple}
\end{equation}
As each $X ^\mu$ operator implies one relation among $\{ X \}$ operators, the number of $\{ X \}$ operators taking all relations into account is the dimension of $\{ X \}$ minus the dimension of $\{ X ^\mu\}$ space, exactly whats accomplished by the $\mathcal O(D^0)$ and $\mathcal O(D)$ terms in Eq.~\eqref{expansion of P}.

The $1/P$ factor doesn't stop at $\mathcal O(D)$ because the IBP relations defined by Eq.~\eqref{eq:ibpsimple} are not always independent. To correct for this, higher order corrections need to be taken into consideration. For example, if an operator in $\{ X ^\mu\}$ can be expressed as $\partial_\nu  X ^{[\mu\nu]}$, where the $[\cdot\cdot\cdot]$ denotes antisymmetrization, then $\partial_\mu\partial_\nu \{ X ^{[\mu\nu]}\}=0$ identically. By the logic above, this zero means each $\{ X ^{[\mu\nu]}\}$ operator implies a linear dependent relation among the previous $\partial_\mu \{ X ^\mu\}$ equations:
\begin{equation}
\partial_\nu\partial_\mu X^{[\mu\nu]}=\sum_mb_m(\partial_\mu X^\mu_m)=0.
\end{equation} 

Iterating, we see that the last two terms, which represent operator spaces $\{ X ^{[\mu\nu\rho]}\}$ and $\{ X ^{[\mu\nu\rho\sigma]}\}$ correct the $\mathcal O(D^2)$ and $\mathcal O(D^3)$ terms respectively. The expansion terminates at $D^4$ because in four dimensions we cannot form a non-trivial operator with five or more totally antisymmetric indices. Therefore no space can correct $\{ X ^{[\mu\nu\rho\sigma]}\}$, and the series ends\footnote{One can understand the termination of the series by realizing that any total derivative is itself a closed but not exact d-forms \cite{Henning:2015daa,Henning:2017fpj}, where d is the dimension of spacetime. In 4 dimensions, one can at most has a 4-form, whose basis is given by $d\omega^\mu\wedge d\omega^\nu\wedge d\omega^\rho\wedge d\omega^\sigma$, with a coefficient carrying antisymmetic indices among $\mu,\nu,\rho,\sigma$. }. 

Putting things together, the number of independent operators modulo IBP in the non-supersymmetric case is given by
\begin{equation}\label{smeft num}
\# \ operators \ including \ IBP =\# \{\mathcal{O}\}-\# \{X^\mu\} +\# \{X^{[\mu\nu]}\} -\# \{X^{[\mu\nu\rho]}\} +\# \{X^{[\mu\nu\rho\sigma]}\}.
\end{equation}

The above understanding of the $\frac{1}{P(D,\alpha,\beta)}$ factor sheds light on how to find similar factors in more general cases to remove IBP relations. For this purpose, we first give a definition of what is a correction and then apply it in the non-supersymmetric case to reproduce the $\frac{1}{P(D,\alpha,\beta)}$. Starting with a space $\mathcal{O}$, we define the zeroth order equivalence relations on $\mathcal{O}$ as follows:
\begin{equation}\label{IBP definition}
\mathcal{o}_1\sim \mathcal{o}_2+\sum \mathcal{I}_i s_i,\ \ \ \ \mathcal{o}_i\in \mathcal{O},s_i\in S^0_i
\end{equation}
where $\mathcal{I}_i$ are maps that take elements from $S^0_i$ to $\mathcal{O}$ and the sum runs over all possible $S^0_i$  and the dimension of each $\mathcal S^0_i$. The upper index $0$ indicates that this is the zeroth-order correction. The IBP relation for a non-supersymmetric theory fits right into this general definition if we identify $\mathcal O$ as the space $\{ X \}$, $\mathcal S^0$ as the space $\{ X ^\mu\}$, and $\mathcal I $ is $\partial_\mu$,
\begin{align}
\mathcal O_i \sim \mathcal O_j + \sum_n \partial_\mu \mathcal O^\mu_n, \quad \mathcal O_i, O_j \in \{ X \}, \mathcal O^\mu_n \in \{ X ^\mu\}.
\end{align}
For non-supersymmetric theories, there is only one class, or branch, of corrections, so there is no $i$ index on $S^0$, however for more general setups there may be multiple $S_i^0$. 

Next, we identify the space $\mathcal{S}^1_j$, along with maps $\mathcal{T}^1_{ij}:\mathcal{S}_j^1\rightarrow S^0_i$. We call $\mathcal{S}^1_j$ the first order correction space if all elements in $\mathcal{S}^1_j$ satisfy the following conditions:
\begin{equation}
\label{eq:mapcriteria}
\mathcal{T}^1_{ij}s_j\neq0,\ \ and\ \ \mathcal{I}_i\mathcal{T}^1_{ij}s_j=0,\ (no\  sums\  over\  i),\forall s_j\in \mathcal{S}^1_j
\end{equation}
From the definition, we see the superscript indicates the order of the correction (1, here), while the subscript $j,i$  respectively label which of the $S^1$ and $S^0$ spaces are connected with the map.
For a non-supersymmetric theory, again there is only one $S^1$ space -- the operator set $\{ X ^{[\mu\nu]}\}$ -- thus the only map, $\mathcal{T}^1_{11} = \partial_\nu$,  has $i = j = 1$. Clearly, all operators in $\{ X ^{[\mu\nu]}\}$ satisfy
\begin{align}
& \mathcal T^1_{11} s = \partial_\nu X^{[\mu\nu]} \ne 0 \nonumber \\
&  \mathcal I_1 \mathcal T^1_{11} s = \partial_\mu\partial_\nu X^{[\mu\nu]} = 0
\end{align}

Higher-order corrections are defined in a similar way. A space $\mathcal{S}_j^n$ is called the nth-order correction to $\mathcal{O}$ if there exist maps $\mathcal{T}^n_{ij}:S^n_j\rightarrow S_i^{(n-1)}$, such that:
\begin{equation}
\label{eq:nthorder}
\mathcal{T}_{ij}^ns_j\neq0,\ \ and\ \ \mathcal{T}_{ki}^{n-1}\mathcal{T}^n_{ij}s_j=0,\forall s_j\in S_j^n,\forall k,
\end{equation}
and is denoted as $\mathcal{S}^n_j(\{\mathcal{S}^{n-1}_i\}\rightarrow \{\mathcal{S}^{n-2}_i\}),n\geq2.$ This notation allows us to keep track of all corrections and maps, such that we can easily prove whether a given space (or spaces) and related maps satisfy the definition.  In our non-supersymmetric example, it is easy to see that the spaces $S_1^2=\{X^{[\mu\nu\rho]}\},\ S_1^3=\{X^{[\mu\nu\rho\sigma]}\}$ and maps $\mathcal{T}_{11}^2=\partial_\rho,\ \mathcal{T}_{11}^3=\partial_\sigma$ satisfy the criteria. 

We can use diagrams to keep track of the corrections and spaces. Starting from the left, we place the space $\mathcal O$. Next comes $\mathcal S^0_i$, with arrows pointing from $\mathcal S^0_i$ to $\mathcal O$ indicating the maps $\mathcal I_i$. The second column is $S^1_j$, with arrows from $S^1_j$ to the $S^0_i$ representing the maps $T^1_{ij}$. Next comes $S^2_j$ with its affiliated maps, then $S^3_j$, and so on. For non-supersymmetric theories, the diagram is shown below in Fig.~\ref{treesmeft}. There is only one correction space at each order (one $S^0$, one $S^1$, etc.), so the correction diagram is a single line; the leftmost space is $\{X\}$, and the diagram ends with $S^3 = \{X^{[\mu\nu\rho\sigma]}\}$.

\begin{figure}[h!]
\begin{center}
\includegraphics[scale=0.45]{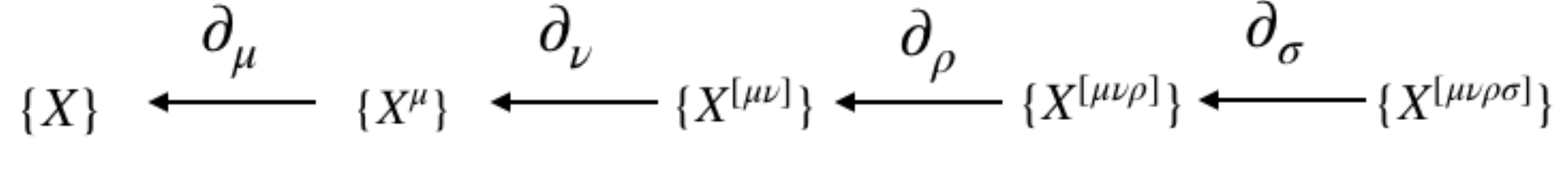}
\end{center}
\caption{This is the diagram in the non-supersymmetric case. It terminates at order $\mathcal O(D^4)$, as explained in the text. The related maps are given by $\partial_\mu$.}\label{treesmeft}
\end{figure}

Having defined the corrections, we are now able to calculate the number of independent operators:
\begin{equation} \label{num}
\# \ of \ independent\  operators=\# \{\mathcal{O}\}-\# \sum\{\mathcal{S}^0_i\}+\# \sum\{\mathcal{S}^1_i\}-\# \sum\{\mathcal{S}^2_i\}+\cdot\cdot\cdot
\end{equation} where the $\# \{X\}$ represents the number of operators in $\{X\}$ space.
The series of corrections may terminate at some fixed order, as in the non-supersymmetric case,  or it may continue infinitely. To execute this counting within the Hilbert series, each of the $S^i_j$ need to be dressed up with the appropriate Lorentz group characters -- so the right spaces are projected out by character orthonormality -- and multiplied by spurions representing the maps $\mathcal I, \mathcal T^n_{ij}$. In non-supersymmetric theories, the map spurions are all just $D$, and the character/spurion dressed version of Eq.~\eqref{num} reproduces $1/P$. 

There is a subtlety that we should mention. For non-supersymmetric theories the Hilbert series can be written as $\mathcal H = \mathcal H_0 + \Delta\mathcal H$~\cite{Henning:2017fpj}. The procedure described above -- plethystic exponential, conformal characters, and $1/P$ factor -- reproduces $\mathcal H_0$. The $\Delta H$ pieces is a correction stemming from the non-orthonormality of the characters for short representations of the conformal group under the Haar measure for the $SO(4)\times SO(2)$ (maximal compact subgroup of the conformal group). For scalars, spinors and field strengths in four dimensions, $\Delta \mathcal H$ only includes terms of dimension four or less. So, while it is needed for full operator basis, $\Delta \mathcal H$ plays no role if our interest is counting higher dimensional operators. Our approach for supersymmetric theories may also generate contributions to $\Delta \mathcal H$,  however, as in the non-supersymmetric scenario, $\Delta \mathcal H$ will only include operators with mass dimension $\le$ four. As such, we will ignore $\Delta \mathcal H$ for the remainder of this work, focusing on the (mass dimension $\ge 4 $) terms contained in $\mathcal H_0$.

%Moreover our method calculates the number of independent operators with a particular number of superfield and superderivatives, it does not (yet) generate the complete set of independent operators.

Now we have everything we need to study the more complicated supersymmetric case. As we will see in the next section, there are two independent IBP relations in supersymmetry, e.g. two spaces $S^0_i$, and the maps $\mathcal T_{ij}$ take on a more complicated form. The net result is a more interesting and subtle correction structure. 

\subsection{IBP relations in $N=1$ supersymmetry}\label{IBP in SUSY}
In a ($N= 1$) supersymmetric theory, there are three possibilities two K\"ahler terms can differ by a total derivative -- the IBP relations,
\begin{equation}
\begin{split}
&K=K'+\partial_\alpha X^\alpha\\
&K=K'+\partial_{\dot{\alpha}} X^{\dot{\alpha}}\\
&K=K'+\partial_\mu X^\mu.\\
\end{split}
\label{sIBP1}
\end{equation}
However, only two of these are independent. Use the defining anti-commutation relations between the two superderivatives $\{\partial_\alpha,\partial_{\dot{\alpha}}\}=2i\sigma^{\mu}_{\alpha\dot{\alpha}}\partial_\mu\equiv\partial_{\alpha\dot{\alpha}}$, we can rewrite the third relation as:
\begin{equation}
K\sim K'+\partial_\alpha(\partial_{\dot{\alpha}}X^{\alpha\dot{\alpha}})+\partial_{\dot{\alpha}}(\partial_\alpha X^{\alpha\dot{\alpha}})
\end{equation}
which is a linear combination of the first two equations. Therefore there are only two independent possibilities in $N=1$ supersymmetry and we choose the first two to be the IBP relations. The above also means we only need two of $\partial_\alpha, \partial_{\dot\alpha}, \partial_\mu$ to build operators. Following our choice for IBP relations, we'll keep $\partial_\alpha$ and $\partial_{\dot\alpha}$; roughly speaking, the reader looking to spot factors of $\partial_\mu$ should look for combinations $\partial_\alpha \partial_{\dot\alpha}$ or $ \partial_{\dot\alpha}\partial_\alpha$ (exactly which depends on whether the object acted on by the derivatives is chiral, antichiral, or neither).

When two K\"ahler terms differ by a total derivative, they will give the same action once integrated over the superspace. For example, in the first case,
\begin{equation}
\begin{split}
\int d^4x\, d^4\theta K&=\int d^4x\,d^4\theta (K'+\partial_\alpha X^\alpha)\\
&=\int d^4x\, d^4\theta K'+\int d^4x\,d^4\theta \partial_\alpha X^\alpha\\
&=\int d^4x\, d^4\theta K'+\int d^4x\,d^2\overline{\theta} \partial^3_\alpha X^\alpha\\
&=\int d^4x\, d^4\theta K'
\end{split}
\label{Kterm}
\end{equation}
where in the third line the integration goes from the full superspace to half superspace, and in the fourth line the second term vanishes because $\partial^3_\alpha=0$. The fact that $K$ and $K'$ give the same action means that we only need to take into account of one of them when we form the Lagrangian.

From now on, we will use a slightly different notation to label different spaces. Let $p'$ and $q'$ represent the number of $\partial_\alpha$ and $\partial_{\dot{\alpha}}$ in each operator space, we define the space $\{X^{\alpha_1\alpha_2\cdot\cdot\cdot\dot{\alpha}_1\dot{\alpha}_2\cdot\cdot\cdot}\}^{p,q}$ to be the space spanned by the basis determined by the Hilbert Series, where $p=m-p'$, $q=n-q'$ and $m,n$ are the number of superderivatives of $\partial_\alpha,\ \partial_{\dot{\alpha}}$ respectively in $\mathcal{O}$, the operator space we are interested in; see Table \ref{table} for details.  The $X^{\alpha_1\alpha_2\cdot\cdot\cdot\dot{\alpha}_1\dot{\alpha}_2\cdot\cdot\cdot}$ indicates the spinorial structure of elements in that space. For example, if we want to study the case $\mathcal{O}(\partial_\alpha^2\partial_{\dot{\alpha}}^2\Phi^2\Phi^{\dagger 2})$, then $(\partial_\alpha\Phi)^2(\partial_{\dot{\alpha}}\Phi^\dagger)^2\in \{X\}^{0,0}=\mathcal{O}$, $(\partial_\alpha\Phi)^2(\Phi^\dagger)^2\in \{X\}^{0,2}$, $\Phi^2\Phi^{\dagger 2}\in \{X\}^{2,2}$, etc. This notation explicitly shows how many superderivatives a space carries, making it easier for us to arrange and order different spaces. At times, we will omit the $\alpha, \dot{\alpha}$ indices in $\{X\}$ for brevity, though they can be reconstructed knowing $p$ and $q$. 

Equations~\eqref{Kterm} and \eqref{sIBP1} motivate the following IBP equivalence relation
\begin{align}
\mathcal{o}_1\sim\mathcal{o}_2+\sum \partial_\alpha X^\alpha+\sum \partial_{\dot{\alpha}} X^{\dot{\alpha}}, \mathcal{o}_i\in\mathcal{O},
\label{susyIBP}
\end{align}
which fits into the zeroth order correction master formula Eq.~\eqref{IBP definition} if we define two correction spaces $S^0_1 =  \{X^{\alpha}\}^{1,0}, S^0_2 = \{X^{\dot{\alpha}}\}^{1,0}$ with corresponding maps $\mathcal I_1 = \partial_\alpha, \mathcal I_2 = \partial_{\dot{\alpha}}$.

Using \eqref{eq:mapcriteria} and \eqref{eq:nthorder}, we can fill out the entire diagram of higher order corrections spaces and maps. The result is shown below in Fig \ref{tree}.   Arrows point from the correction space to the space they correct, i.e. from $S^n$ to $S^{n-1}$. The zeroth order corrections lie in the second column, $S^0_1 = \{X\}^{1,0}$ and $S^0_2 = \{X\}^{1,0}$, with maps $\partial_\alpha$ and $\partial_{\dot{\alpha}}$ connecting them to $\mathcal O = \{X\}^{0,0}$, as expected from Eq.~\eqref{susyIBP}. The higher corrections are naturally divided into six `branches', three of which are oriented in the same direction as the $\mathcal I_1 = \partial_\alpha$ zeroth order map and three which are oriented along the $\mathcal I_2 = \partial_{\dot{\alpha}}$ map direction. For simplicity, we'll refer to these two groups as the `$\partial_\alpha$' and `$\partial_{\dot{\alpha}}$' directions. They are symmetric under the change $\alpha\leftrightarrow\dot{\alpha}$. The expressions for the higher order maps, $l_n$ and $\overline{l}_n$, are more complicated and will be given shortly. In addition to the multiple branches, another difference between the supersymmetric and non-supersymmetric cases is that the branches in supersymmetric theories do not terminate. 

\begin{figure}[h!]
\begin{center}
\includegraphics[scale=0.35]{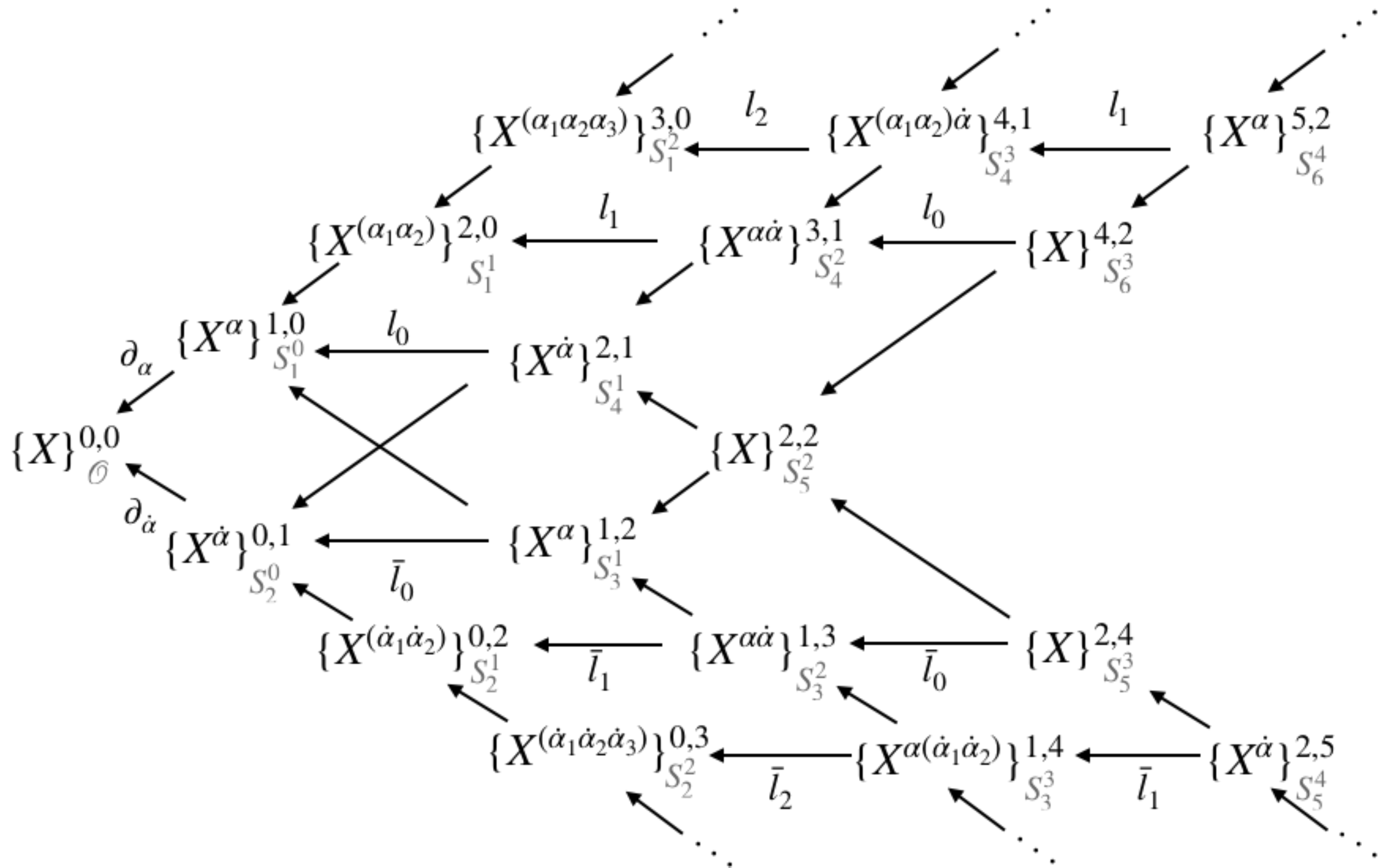}
\end{center}
\caption{This is the tree diagram to illustrate how corrections work. Arrows point from the correction spaces to the spaces they correct. Maps above the arrows represent the maps: $\partial_\alpha, \partial_{\dot{\alpha}}$ for the zeroth order maps, and $l_n, \bar{l}_{n}$ for higher orders. See the text for the explicit form of $l_n, \bar{l}_n$.}
\label{tree}
\end{figure}

Notice that the number of spaces increases with the correction order: there are two $S^0$, four $S^1$, five $S^2$, etc. 
To check if these spaces and the maps that connect them satisfy our map criteria Eq.~\eqref{eq:mapcriteria}, we need to choose a labeling scheme. We choose $( S^0_1 = \{X\}^{1,0}, S^0_2 = \{X\}^{0,1})$ for the zeroth order corrections, $(S^1_1 = \{X\}^{2,0}, S^1_2 = \{X\}^{0,2}, S^1_3 = \{X\}^{2,1}, S^1_4 = \{X\}^{1,2} )$ for the first order,  and $(S^2_1 = \{X\}^{3,0}, S^2_2 = \{X\}^{0,3}, S^2_3 = \{X\}^{3,1}, S^2_4 = \{X\}^{1,3}, S^2_5 = \{X\}^{2,3} \}$ for second order\footnote{For a different labeling, the only change would be in the $\mathcal T_{ij}$ indices. While the space of corrections in infinite, we have only listed the explicit labeling scheme for the spaces needed to prove the relations in the text.}. Each space has both an $\{X\}$ name, which tells us the derivative content of its operators, and an $S$ name, which orients the map with respect to the original space $\mathcal O$, distinguishes between equidistant maps, and most compactly expresses the IBP relations. Table~\ref{table} below shows both names of the spaces we are interested in, along with their Lorentz group representation and characters.

\begin{table}[h!]
\begin{center}
\begin{tabular}{ |c|c|c|c| }
 \hline
 \multicolumn{4}{|c|}{Corrections} \\
 \hline
  $S_i^n$ &  $X^{p,q}$&Representation&Character\\
 \hline
  $ $    &  $\{X\}^{0,0}$&$(0,0)$   &1\\
   $S_1^{n-1} (n\geq1)$ & $\{{X^{(\alpha_1\alpha_2\cdot\cdot\cdot\alpha_n)}}\}^{n,0}$&$(\frac{n}{2},0)$&$\frac{\sin((n+1)\Omega_x)}{\sin\Omega_x}$\\
 $S_2^{n-1} (n\geq1)$ & $\{{X^{(\dot{\alpha}_1\dot{\alpha}_2\cdot\cdot\cdot\dot{\alpha}_n)}}\}^{0,n}$&$(0,\frac{n}{2})$ &$\frac{\sin((n+1)\Omega_y)}{\sin\Omega_y}$\\
 $S_3^1$ & $\{{X^\alpha}\}^{1,2}$&$(\frac{1}{2},0)$    &$x$\\
$S_4^1$ & $\{{X^{\dot{\alpha}}}\}^{2,1}$&$(0,\frac{1}{2})$    &$y$\\
 $S_3^n (n\geq 2)$ & $\{{X^{\alpha(\dot{\alpha}_1\dot{\alpha}_2\cdot\cdot\cdot\dot{\alpha}_{n-1})}}\}^{1,n+1}$&$(\frac{1}{2},\frac{n-1}{2})$    &$x\frac{\sin(n\Omega_y)}{\sin\Omega_y}$\\
    $S_4^n (n\geq2)$  & $\{{X^{\dot{\alpha}(\alpha_1\alpha_2\cdot\cdot\cdot\alpha_{n-1})}}\}^{n+1,1}$&$(\frac{n-1}{2},\frac{1}{2})$&$y\frac{\sin(n\Omega_x)}{\sin\Omega_x}$\\
$S_5^{2}$     & $\{X\}^{2,2}$&$(0,0)$   & 1\\
 $S_5^{3}$    & $\{X\}^{2,4}$&$(0,0)$   &1\\
 $S_6^{3}$     & $\{X\}^{4,2}$&$(0,0)$   &1\\
  $S_5^{n} (n\geq4)$ & $\{{X^{(\dot{\alpha}_1\dot{\alpha}_2\cdot\cdot\cdot\dot{\alpha}_{n-3})}}\}^{2,n+1}$& $(\frac{n-3}{2},0)$&$\frac{\sin((n-2)\Omega_x)}{\sin\Omega_x}$\\
 $S_6^{n} (n\geq4)$ & $\{{{X}^{(\dot{\alpha}_1\dot{\alpha}_2\cdot\cdot\cdot\dot{\alpha}_{n-3})}}\}^{n+1,2}$&$(0,\frac{n-3}{2})$ &$\frac{\sin((n-2)\Omega_y)}{\sin\Omega_y}$\\
 \hline
\end{tabular}
\end{center}
\caption{This table summarizes the representation of each space and the corresponding character. The translation from $X$ notation to $S$ notation is also provided. Here, $x$ and $y$ are defined as $x=\alpha+\frac{1}{\alpha}$ and $y=\beta+\frac{1}{\beta}$, where $\alpha, \beta$ are the $SU(2)_L, SU(2)_R$ group parameters, and $\Omega_{x,y}$ are defined by $x,y\equiv 2\cos(\Omega_{x,y})$.}
\label{table}
\end{table}

The higher order corrections are best understood moving along the diagonal branches, rather than thinking in columns. In the following sections we will study the three `$\partial_\alpha$ branches' in detail. 

\subsubsection{First branch}\label{first branch}

\begin{figure}
\begin{center}
\includegraphics[scale=0.4]{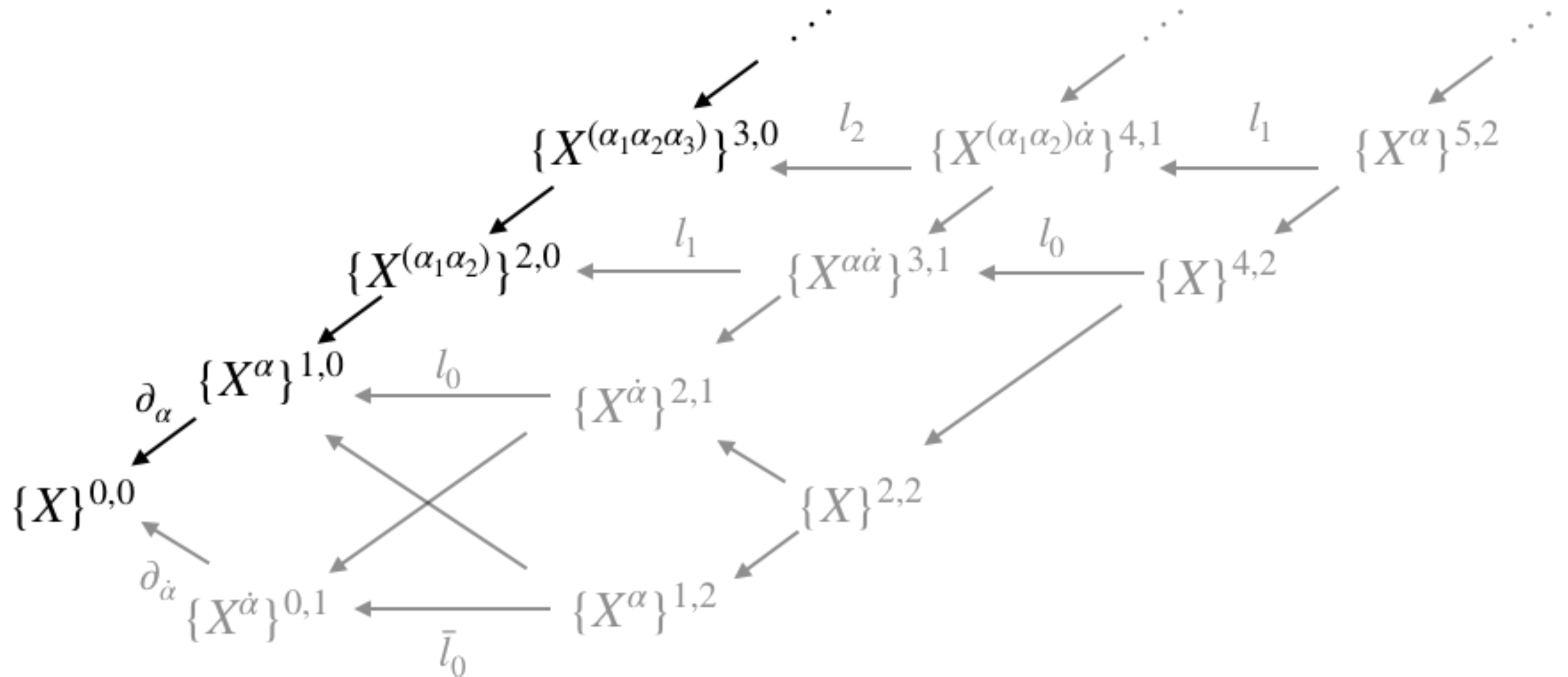}
\end{center}
\caption{The bold text shows the first IBP  branch for supersymmetric theories. It extends to infinity, unlike what happens in non-supersymmetric theories; $\{X^{(\alpha_1\alpha_2\cdot\cdot\cdot\alpha_n)}\}^{n,0}$ corrects $\{X^{(\alpha_1\alpha_2\cdot\cdot\cdot\alpha_{n-1})}\}^{n-1,0}$ and the map is given by $\partial_\alpha$.}\label{tree1}
\end{figure}

The first branch we study is shown in Fig.~\ref{tree1}, continues along the $\partial_\alpha$ direction and contains the correction spaces $\{X\}^{2,0}, \{X\}^{3,0},\cdots$. This branch, and its complex conjugate $\partial_{\dot{\alpha}}$ branch, are the most similar to the non-supersymmetric case. We will use that similarity to intuit the result, then prove that all maps and spaces satisfy the required criteria (Eq.~\eqref{eq:mapcriteria}).

Recall the counting formula given in \eqref{smeft num}. There, each space carries totally antisymmetric indices, and as a result vanishes when acting on two (commuting) partial derivatives. In four dimensions, the maximal number of fully antisymmetric indices is four, and therefore the series terminates at $\{X^{[\mu\nu\rho\sigma]}\}$. Let's try the same trick in the supersymmetry case. As all superderivatives are fermionic, i.e. anti-commuting, acting two of them on an operator that carries totally {\it symmetric} indices will vanish identically, i.e. $\partial_\alpha\partial_\beta X^{(\alpha\beta)} = 0$. Each vanishing combinations implies a relation among operators and is thus a correction, just as in the non-supersymmetric case. However, the spacetime dimension places no restriction on operators with symmetric $SU(2)$ (Lorentz) indices an operator can carry. Therefore, in the supersymmetry case, this type of correction does not terminate at a fixed derivative order. In parallel with \eqref{smeft num}, we expect the counting formula for supersymmetric theories coming from this branch is given by:
\begin{equation}
\# \ comes\ from\ the\ frist\ branch=-\# \{X^{\alpha_1}\} +\# \{X^{(\alpha_1\alpha_2)}\} -\cdot\cdot\cdot+(-1)^n\# \{X^{(\alpha_1\alpha_2\cdot\cdot\cdot\alpha_{n})}\}\cdot\cdot\cdot
\end{equation}
Now we will prove this formula using the definition and give one example of the existence of higher order corrections.

It is clear that we can get $\# \{X\}^{1,0}$ relations among $\{X\}^{0,0}$ operators from acting $\partial_\alpha$ on each term in $\{X\}^{1,0}$. However, just as in the non-supersymmetric case, these relations may not be independent. For example, acting $\partial_\alpha$ on $\partial_\beta\Phi X^{(\alpha\beta)}$ and $\Phi \partial_\beta X^{(\alpha\beta)}$ gives the same relations, even though they come from different terms. The reason is because of the fact that $\partial_\alpha\partial_\beta(\Phi X^{(\alpha\beta)})\sim 0$ identically. It's not difficult to extend this to general $n$. We claim that $S_1^{n}=\{X^{(\alpha_1\alpha_2\cdot\cdot\cdot\alpha_{n+1})}\}$ corrects $S_1^{n-1}=\{X^{(\alpha_1\alpha_2\cdot\cdot\cdot\alpha_{n})}\}$, where $(\cdot\cdot\cdot)$ represents fully symmetrization of the indices. Any element $s_1^n$ in $S_1^n$ transforms under $(\frac{n+1}{2},0)$ and the proof is straightforward as follows: 
\begin{equation}
\mathcal{T}_{11}^n s^n_1\neq0,\ and\ \mathcal{T}_{11}^{n-1}\mathcal{T}_{11}^n s^n_1=\partial_{\alpha_1} \partial_{\alpha_2}X^{(\alpha_1\alpha_2\cdot\cdot\cdot\alpha_{n+1})}=0.
\end{equation}
The second equation vanishes because of the antisymmetric property of superderivatives. 

\subsubsection{Second branch}
\label{second branch}

\begin{figure}
\begin{center}
\includegraphics[scale=0.4]{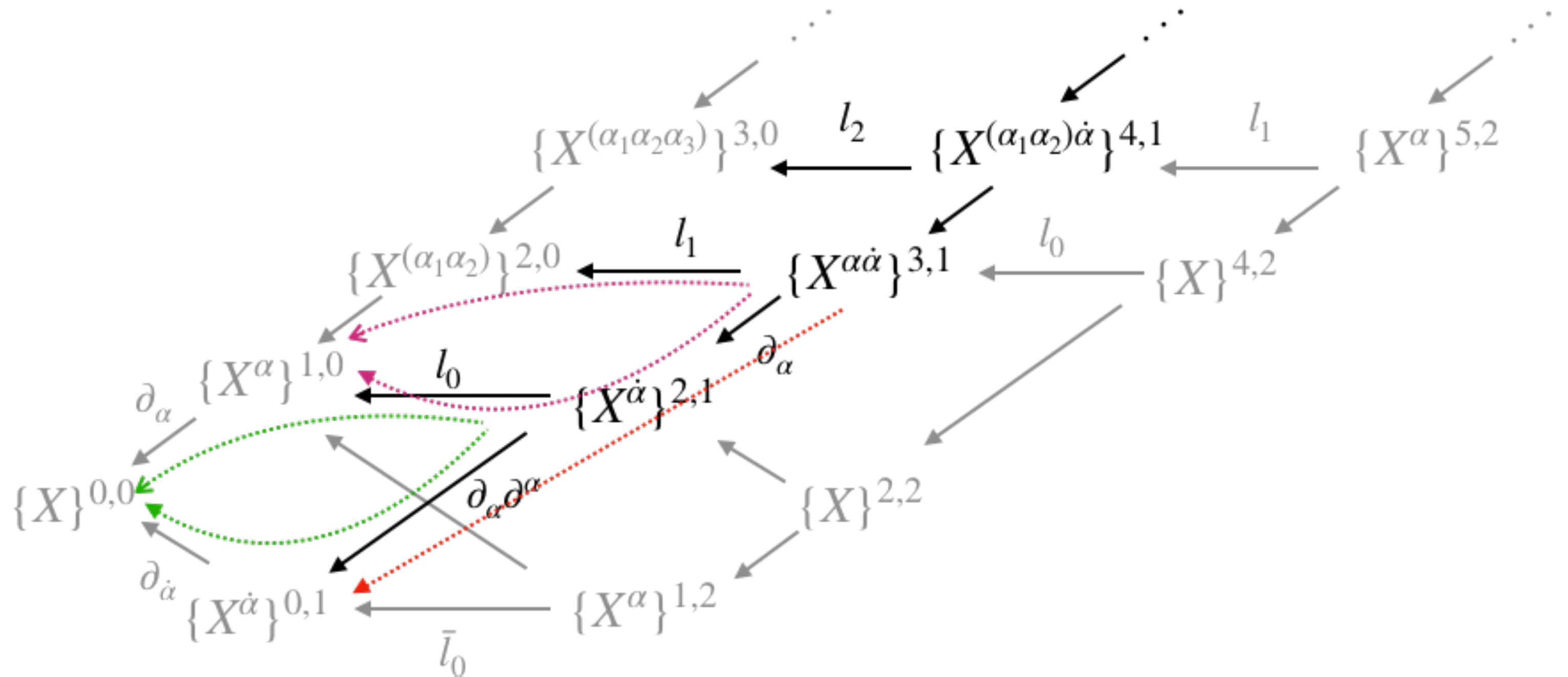}
\end{center}
\caption{$\{X^{(\alpha_1\alpha_2\cdot\cdot\cdot\alpha_{n-2})\dot{\alpha}}\}^{n,1}\ (n\geq2)$ represents the second branch and each term corrects two spaces as can be read from the diagram. The green loop represents the first example, while the purple loop and red straight line illustrate the second example, as explained in the text. The maps $l_n$ are given in the text.}\label{tree2}
\end{figure}

In this section, we will study the second branch, shown in Fig.~\ref{tree2}. The second branch is a new feature in supersymmetry, arising from the fact that the theory has two superderivatives as well as one ordinary partial derivative. Although they are not independent, two of them survive after removing one of them using the defining anticommutation relation \eqref{anti relation}. To see how the second branch comes about, let's look at two examples at low dimensions, and then we give the general result and proof for arbitrary dimensions.  

Suppose we want to build the operator basis for $\{X\}^{0,0}=\mathcal{O}(\partial_\alpha^2\partial_{\dot{\alpha}}^2\Phi^2\Phi^\dagger)$. Then we know that we can get the following IBP relations from $\{X\}^{1,0}=\mathcal{O}(\partial_\alpha\partial_{\dot{\alpha}}^2\Phi^2\Phi^\dagger)$ and $\{X\}^{0,1}=\mathcal{O}(\partial_\alpha^2\partial_{\dot{\alpha}}\Phi^2\Phi^\dagger)$:
\begin{subequations}
\begin{align}
&\partial_{\dot{\alpha}}(\partial_\alpha\Phi\partial^\alpha\Phi\partial^{\dot{\alpha}}\Phi^\dagger)\sim 2\partial_{\dot{\alpha}}\partial_\alpha\Phi\partial^\alpha\Phi\partial^{\dot{\alpha}}\Phi^\dagger\sim 0,\\
&\partial_{\dot{\alpha}}(\partial_\alpha\Phi\Phi\partial^\alpha\partial^{\dot{\alpha}}\Phi^\dagger)\sim \partial_{\dot{\alpha}}\partial_\alpha\Phi\Phi\partial^\alpha\partial^{\dot{\alpha}}\Phi^\dagger\sim 0,\\
&\partial_\alpha(\partial^{\dot{\alpha}}\partial^\alpha\Phi\Phi \partial_{\dot{\alpha}}\Phi^\dagger)\sim \partial^{\dot{\alpha}}\partial^\alpha\Phi\partial_\alpha\Phi \partial_{\dot{\alpha}}\Phi^\dagger+\partial^{\dot{\alpha}}\partial^\alpha\Phi\Phi \partial_\alpha\partial_{\dot{\alpha}}\Phi^\dagger\sim 0.
\end{align}
\end{subequations}
where we have dropped all EOM terms, e.g. $\partial^2_\alpha \Phi$ or $\partial^\alpha\partial_{\dot\alpha}\partial_\alpha\Phi$ when acting with the final derivative; it is also important to remember that $\partial_\alpha, \partial_{\dot\alpha}$ are Grassmann objects with indices raised/lowered with $\epsilon$. It's clear that $(c)=\frac 12 (a)+(b)$, so only two of the above three relations are independent.  The reason for the connection is that $\partial_\alpha\Phi\partial^\alpha\Phi\partial^{\dot{\alpha}}\Phi^\dagger$ and $\partial_\alpha\Phi\Phi\partial^\alpha\partial^{\dot{\alpha}}\Phi^\dagger$ can all be obtained from an operator with even fewer derivatives, $\Phi^2\partial_{\dot{\alpha}}\Phi^\dagger\in \{X\}^{2,1}$. Specifically,  $\partial_\alpha\Phi\partial^\alpha\Phi\partial^{\dot{\alpha}}\Phi^\dagger$ and $\partial_\alpha\Phi\Phi\partial^\alpha\partial^{\dot{\alpha}}\Phi^\dagger$ both are generated by applying $(\partial_\alpha)^2$ to $\Phi^2\partial_{\dot{\alpha}}\Phi^\dagger$, while $\partial^{\dot{\alpha}}\partial^\alpha\Phi\Phi \partial_{\dot{\alpha}}\Phi^\dagger$ is generated from $(2\partial_{\dot\alpha}\partial^{\alpha}+\partial^{\alpha}\partial_{\dot\alpha})\Phi^2\partial_{\dot{\alpha}}\Phi^\dagger \equiv l_0 (\Phi^2\partial_{\dot{\alpha}}\Phi^\dagger)$. As the three `daughters' $\partial_\alpha\Phi\partial^\alpha\Phi\partial^{\dot{\alpha}}\Phi^\dagger$, $\partial_\alpha\Phi\Phi\partial^\alpha\partial^{\dot{\alpha}}\Phi^\dagger$ and $\partial^{\dot{\alpha}}\partial^\alpha\Phi\Phi \partial_{\dot{\alpha}}\Phi^\dagger$ share a single `mother', the relations they imply ($(a)$ through $(c)$) are interconnected.

Let us verify that these maps do indeed satisfy Eq.~\eqref{eq:mapcriteria}. This will better illustrate how to unpack Eq.~\eqref{eq:mapcriteria}, as well as provide some more physical insight into the relations. Referring to Table~\ref{table}, $S^1_3=\{X\}^{2,1} $, $S^0_1 = \{X\}^{1,0}, S^0_2 = \{X\}^{0,1}$, and we have already identified the zeroth order correction maps $\mathcal I_1 = \partial_\alpha$, $\mathcal I_2 = \partial_{\dot\alpha}$. What remains are the maps taking us from $\{X\}^{2,1}$ to $\{X\}^{1,0}$ and $\{X\}^{0,1}$. In the $S$ notation, the maps take $S^1_3$ to $S^0_1$ and $S^0_2$, so $\mathcal T^1_{13}, \mathcal T^1_{23}$. From the discussion above, we see $\mathcal T^1_{13} = l_0 $ and $\mathcal T^1_{23} = \partial^2_\alpha$, where $l_0$ is defined as:
\begin{equation}\label{l_0}
(l_0)^\alpha_{\dot{\alpha}}=2\partial_{\dot\alpha}\partial^{\alpha}+\partial^{\alpha}\partial_{\dot\alpha}.
\end{equation}
For future reference we also introduce the definition of $l_1$ acting on $X^{\alpha\dot{\alpha}}$ to give $X^{(\alpha_1\alpha_2)}$:
\begin{equation}
(l_1)_{\alpha\dot{\alpha}}^{\alpha_1\alpha_2}X^{\alpha\dot{\alpha}}=-(\partial_{\dot{\alpha}}\partial_\sigma+\frac{2}{3}\partial_\sigma\partial_{\dot{\alpha}})(\epsilon^{\alpha_1\beta}\epsilon^{\sigma\alpha_2}+\epsilon^{\alpha_2\beta}\epsilon^{\sigma\alpha_1}) \epsilon_{\beta\alpha}X^{\alpha\dot{\alpha}}
\end{equation}
 For these corrections, Eq.~\eqref{eq:mapcriteria} becomes:
\begin{align}\label{second branch proof}
& (\mathcal I_1\,\mathcal T^1_{13} + \mathcal I_2 \mathcal T^1_{23})S^1_3 = (\partial_\alpha l_0 + \partial_{\dot\alpha}\partial^2_{\alpha})S^1_3 \nonumber \\
& = 2\,\partial_\alpha \partial_{\dot\alpha}\partial^\alpha + \partial_\alpha\partial^\alpha \partial_{\dot\alpha} + \partial_{\dot\alpha}\partial_\alpha\partial^\alpha \nonumber \\
& = 2\,\partial_\alpha \partial_{\dot\alpha}\partial^\alpha  - (\{\partial^\alpha \{\partial_\alpha, \partial_{\dot\alpha} \}  - \partial^\alpha\partial_{\dot\alpha}\partial_\alpha) +  \{ \partial_{\dot\alpha}, \partial_\alpha \}\partial^\alpha - \partial_\alpha\partial_{\dot\alpha}\partial^\alpha \nonumber \\
& =  2\,\partial_\alpha \partial_{\dot\alpha}\partial^\alpha  - \partial^\alpha \{\partial_\alpha, \partial_{\dot\alpha} \}  - \partial_\alpha\partial_{\dot\alpha}\partial^\alpha +  \{ \partial_{\dot\alpha}, \partial_\alpha \}\partial^\alpha - \partial_\alpha\partial_{\dot\alpha}\partial^\alpha \nonumber \\
& = [\{ \partial_{\dot\alpha}, \partial_\alpha \},\partial^\alpha] = 0
\end{align}
where the sign flips in the second and third lines come from changing the order of raised/lowered indices with $\epsilon^{\alpha\beta}$, etc. and the last equality comes from Eq.~\eqref{eq:useful}. Using $\{ \partial_{\dot\alpha}, \partial_\alpha \} \sim \partial_\mu$, we can get some intuition for the physics of the second correction. Starting from an operator in $\{X\}^{2,1}$, we can get to $\{X\}^{0,0}$ either by applying a partial derivative $\partial_\mu$ first and then $\partial_\alpha$ or by applying $\partial_\alpha$ first and then $\partial_\mu$. However, the connection between $\partial_\alpha$, $\partial_{\dot\alpha}$ and $\partial_\mu$ tells us these operations commute and the corrections are not independent.

Tracing through the above steps in Fig.~\ref{tree2} in green, we see the two paths ($\partial_\mu$ then $\partial_\alpha$ or vice versa) form a closed loop. The correspondence of the loops in the diagram with the commutation relation $ [\{ \partial_{\dot\alpha}, \partial_\alpha \},\partial^\alpha]  = 0$ will help us identify other second order corrections directly from Fig.~\ref{tree}.

As a second example, consider the correction from the space $\{X\}^{3,1} = S^2_3$. This space corrects $\{X\}^{2,1} = S^1_3$ and $\{X\}^{2,0} = S^1_1$, both of which correct $\{X\}^{1,0} = \mathcal S^0_1$. Additionally $\{X\}^{2,1} = S^1_3$ also corrects $\{X\}^{0,1} = \mathcal S^0_2$. Using the notation introduced after Eq.~\eqref{eq:nthorder}, we can express this compound correction structure as $S^1_3(S^1_{1,3} \to \mathcal S^0_1, S^1_3 \to \mathcal S^0_2)$. The maps required are $\mathcal T^2_{33}$ (connecting $S^2_3 \to S^1_3$), $\mathcal T^{2}_{13}$ (connecting $S^2_3$ to $S^1_1$), $T^1_{13}$ (connecting $S^1_3 \to S^0_1$), $\mathcal T^1_{11}$ (connecting $S^1_1$ to $\mathcal S^0_1$) and $T^1_{23}$ (connecting $S^1_3 \to S^0_2$). From our previous example, we know $T^1_{23} = \partial^2_\alpha$, $\mathcal T^1_{13} = l_0$ and $\mathcal T^1_{11} = \partial_\alpha$. If we make the identification $\mathcal T^2_{33} = \partial_\alpha$, $\mathcal T^2_{13} = (\partial_{\dot\alpha}\partial^{\alpha}+\frac{2}{3}\partial^{\alpha}\partial_{\dot\alpha} ) \equiv l_1$, the maps satisfy $\mathcal{T}_{ki}^{n-1}\mathcal{T}^n_{ij}s=0,\forall s\in S_j^n$ for all $k$:
\begin{align}
\mathcal T^1_{23}\mathcal T^2_{33} S^2_3 = 0 &\leftrightarrow \partial^2_\alpha \partial_\alpha \{X^{\beta\dot{\beta}} \} = 0 \nonumber \\
(\mathcal T^1_{11}\mathcal T^2_{13} + \mathcal T^1_{13}\mathcal T^2_{33})S^2_3 = 0 & \leftrightarrow (\partial_\alpha l_1 + l_0 \partial_\alpha)\{X^{\beta\dot{\beta}} \} = 0 
\label{eq:secondexample}
\end{align}
The first equation is trivial since $\partial_\alpha^3=0$. The second equation is more subtle, but can be proven as follows:
\begin{align}
\label{eq:secondloop}
&( l_0 \partial_\alpha + \partial_\alpha l_1)\{X^{\beta\dot{\beta}} \}  \nonumber \\
= &(2\partial_{\dot\alpha}\partial^{\beta}+\partial^{\beta}\partial_{\dot\alpha})\partial_\alpha s^{\alpha\dot{\alpha}}-\partial_\alpha (\partial_{\dot\alpha}\partial^{\alpha}+\frac{2}{3}\partial^{\alpha}\partial_{\dot\alpha})s^{\beta\dot{\alpha}}-\partial_\alpha (\partial_{\dot\alpha}\partial^{\beta}+\frac{2}{3}\partial^{\beta}\partial_{\dot\alpha})s^{\alpha\dot{\alpha}} \nonumber \\
= &[(2\partial_\gamma\partial_{\dot{\alpha}}\partial^\beta+2\partial_\gamma\partial^\beta\partial_{\dot{\alpha}}-\partial^\beta\partial_{\dot{\alpha}}\partial_\gamma)-(\partial_\alpha\partial_{\dot\alpha}\partial^{\alpha}+\frac{2}{3}\partial_\alpha\partial^{\alpha}\partial_{\dot\alpha})\delta^\beta_\gamma-(\partial_\gamma\partial_{\dot{\alpha}}\partial^{\beta}+\frac{2}{3}\partial_\gamma\partial^{\beta}\partial_{\dot{\alpha}})]s^{\gamma\dot{\alpha}} \nonumber \\
= &(\partial_\gamma\partial_{\dot{\alpha}}\partial^\beta-\partial_\alpha\partial_{\dot\alpha}\partial^{\alpha}\delta^\beta_\gamma-\partial_\gamma\partial_{\dot{\alpha}}\partial^{\beta})s^{\gamma\dot{\alpha}} \nonumber \\
= &\, 0
\end{align}
In the first line, the $\partial_\alpha l_1$ piece turns into two terms, while $l_0 \partial_\alpha$ does not. This difference comes from the index structure of the intermediate maps. Specifically, $\{X^{(\alpha_1\alpha_2)}\}^{2,0} = S^1_1$ is symmetric in the undotted indices, so we need to symmetrize the indices of $l_1$ acting on $\{X^{\alpha\dot{\alpha}}\}^{3,1}$. The intermediate space on the $l_0 \partial_\alpha$ path, $\{X^{\dot\alpha}\}^{2,1}$, has no such symmetrization requirement. Notice that the commutation of $\partial_\alpha$ and $\partial_{\dot\alpha}$ with $\partial_\mu$, Eq.~\eqref{eq:useful}, makes an appearance again in the third and fourth lines. One can refer to the purple loop in Figure \ref{tree2} for illustration.

In these two examples, we've shown how to use Eq.~\eqref{eq:nthorder} and seen the connection between the corrections and relations among the superderivatives and ordinary derivative. However, we provided the maps $l_0$ and $l_1$.  To understand the structure of $l_0,l_1$ in terms of $\partial_\alpha$ and $\partial_{\dot\alpha}$ and to extrapolate to more general scenarios, let us derive the map $l_n$, defined to map the space $\{X\}^{n+2,1} \to \{X\}^{n+1,0}$  \footnote{Recall, $l_0$ mapped $\{X\}^{2,1} \to \{X\}^{1,0}$ while $l_1$ maps $\{X\}^{3,1} \to \{X\}^{2,0}$.}\footnote{The barred maps $\bar l_0, ... \bar l_n$ can be derived from $l_0, ... \l_n$ by swapping $\partial_\alpha \leftrightarrow \partial_{\dot{\alpha}}$}.

In terms of Lorentz group representations, an element of $\{X\}^{n+2,1}$ lies in the $(\frac n 2,\frac 1 2)$ representation (an operator $X^{\dot{\alpha}}_{(\alpha_1\alpha_2\cdot\cdot\cdot\alpha_n)}$) while an element of $\{X\}^{n+1,0}$ lies in the $(\frac {n+1} 2, 0)$ representation (an operator $X_{(\alpha_1\alpha_2\cdot\cdot\cdot\alpha_{n+1})}$ ). A map between these spaces must contract the $\dot\alpha$ index in $X^{\dot{\alpha}}_{(\alpha_1\alpha_2\cdot\cdot\cdot\alpha_n)}$ and one of the $\alpha_i$ indices, maintaining the symmetry of the remaining $(n-1)$ $\alpha_i$. The most general way of doing this, is by contracting over a dummy set (the $\alpha_i$ below) of indices and symmetrize the free indices (the $\beta_i$ below):
\begin{equation}\label{ln0}
(l_n)_{\dot{\alpha}}^{(\alpha_1\alpha_2\cdot\cdot\cdot\alpha_n)(\tau\beta_1\beta_2\cdot\cdot\cdot\beta_n)}X^{\dot{\alpha}}_{(\alpha_1\alpha_2\cdot\cdot\cdot\alpha_n)}=(a_n\partial_{\dot{\alpha}}\partial_Z+b_n\partial_Z\partial_{\dot{\alpha}})\epsilon^{(Z\alpha_1\alpha_2\cdot\cdot\cdot\alpha_n)(\tau\beta_1\beta_2\cdot\cdot\cdot\beta_n)}X^{\dot{\alpha}}_{(\alpha_1\alpha_2\cdot\cdot\cdot\alpha_n)}
\end{equation}
where $a_n$ and $b_n$ are constants and $\epsilon^{(X\alpha_1\alpha_2\cdot\cdot\cdot\alpha_n)(\tau\beta_1\beta_2\cdot\cdot\cdot\beta_n)}$ is defined to be the fully symmetrization permutations of its indices, e.g. $\epsilon^{(\alpha_1\alpha_2)(\beta_1\beta_2)}=\epsilon^{\alpha_1\beta_1}\epsilon^{\alpha_2\beta_2}+\epsilon^{\alpha_2\beta_1}\epsilon^{\alpha_1\beta_2}$. 
We can solve for $a_n, b_n$ by requiring that $l_n$ satisfy the generalized version of Eq.~\eqref{eq:secondloop},
\begin{align}
(\partial_{\beta_n} (l_n)_{\dot{\alpha}}^{(\alpha_1\alpha_2\cdot\cdot\cdot\alpha_{n-1})(\tau\beta_1\beta_2\cdot\cdot\cdot\beta_n)}+(l_{n-1})_{\dot{\alpha}}^{(\alpha_1\alpha_2\cdot\cdot\cdot\alpha_{n-1})(\tau\beta_1\beta_2\cdot\cdot\cdot\beta_{n-1})}\partial^{\alpha_n})\,X^{\dot{\alpha}}_{(\alpha_1\alpha_2\cdot\cdot\cdot\alpha_n)}=0
\end{align}
The details of this exercise are shown in Appendix~\ref{proof section}, yielding:
\begin{align}
\label{eq:aandb}
a_n=(-1)^n\frac{2}{(n+1)!},\ \ b_n=(-1)^n\frac{2}{(n+2)n!}.
%a_n=b_{n-1}-a_{n-1},\ \ -(n+1)a_n=a_{n-1}
\end{align}
With $l_n$ determined, it is easy to see that the corrections from an arbitrary space $\{X\}^{n,1}, n\ge 4$ on the second branch satisfy the required equations. We used the `loop' relation $(\partial_\alpha l_n + l_{n-1}\partial_\alpha)$ to derive $l_n$, so that is automatically satisfied. The only other relation is $\partial_\alpha \partial_\beta \{X^{(\alpha_1\alpha_2 \cdots \alpha_n)\dot\alpha}\}$ - the analog of the first equation in Eq.~\eqref{eq:secondexample} -- but this vanishes as the Grassman derivatives must be antisymmetrized.

It is natural to ask why the second branch starts from $\{X\}^{2,1}$, not some other spaces. Naively one would imagine that the nearest spaces should give the corrections to $\{X\}^{1,0}$, in this case are $\{X\}^{2,0}$ and $\{X\}^{1,1}$\footnote{ $\{X\}^{1,1}$ is the usual $X^{\mu}$ space in the SMEFT case, and this space is removed because we write partial derivative in terms of the 2 superderivatives. As a result,  $\{X\}^{1,1}$  plays no role when we construct correction spaces and doesn't appear in any diagrams.}. We give a heuristic proof in Appendix \ref{starting space}.

\subsubsection{Third branch}\label{third branch}

%\adam{add in S names?}
In this section, we will study the third branch. From our experience with the second branch, the consistency equations show up graphically as compound maps between two spaces separated by two `steps'. All third branch spaces are connected to three (hence the name) spaces via compound maps.
For example, $\{X\}^{5,2}$ is two steps away from $\{X\}^{2,2}$, $\{X\}^{3,1}$ and $\{X\}^{3,0}$. The connection between $\{X\}^{5,2}$ and $\{X\}^{2,2}$ is two steps along the $\partial_\alpha$ direction, while the connection between $\{X\}^{5,2}$ and $\{X\}^{3,1}$ has the loop form $(\partial_\alpha l_1 + l_0 \partial_\alpha)$ -- both of which are familiar from the second branch and can be shown to satisfy Eq.~\eqref{eq:nthorder} using the same logic as there. The new feature of the third branch is the `horizontal' compound map, which takes $\{X\}^{5,2} \to \{X\}^{3,0}$, or, more generally, $\{X\}^{n,2} \to \{X\}^{n-2,0}$. To prove this satisfies the criteria, we need 
\begin{align}
l_n\,l_{n-1} X^{(\alpha_1 \alpha_2 \cdots \alpha_{n-1})}= 0
\end{align}
The proof of this equation is given in Appendix~\ref{proof section}. We have no freedom in this proof, as  $l_n$ have been fixed by the requirement $(\partial_\alpha l_n + (l_{n-1})\partial_\alpha) = 0$. Interestingly, the final step of the proof involves
\begin{align}
\frac{1}{(n+1)(n+2)}[\{\partial^\alpha,\partial_{\dot{\beta}}\},\{\partial_\alpha,\partial_{\dot{\alpha}}\}]
\end{align}
which vanishes as ordinary partial derivatives commute. The intuitive way to understand this is by noticing that we are acting  a composite antisymmetric operator, and the only choice in this case is $\partial_\mu\partial_\nu-\partial_\nu\partial_\mu$. For example, the composite $l_1l_0$ takes element from $\{X\}^{4,2}$ to $\{X\}^{2,0}$, which effectively takes the representation from $(0,0)$ to $(1,0)$ by acting the composite map $[\partial_\mu,\partial_\nu]$, identically vanishes. The same argument works for higher-order correction spaces.

\begin{figure}
\begin{center}
\includegraphics[scale=0.4]{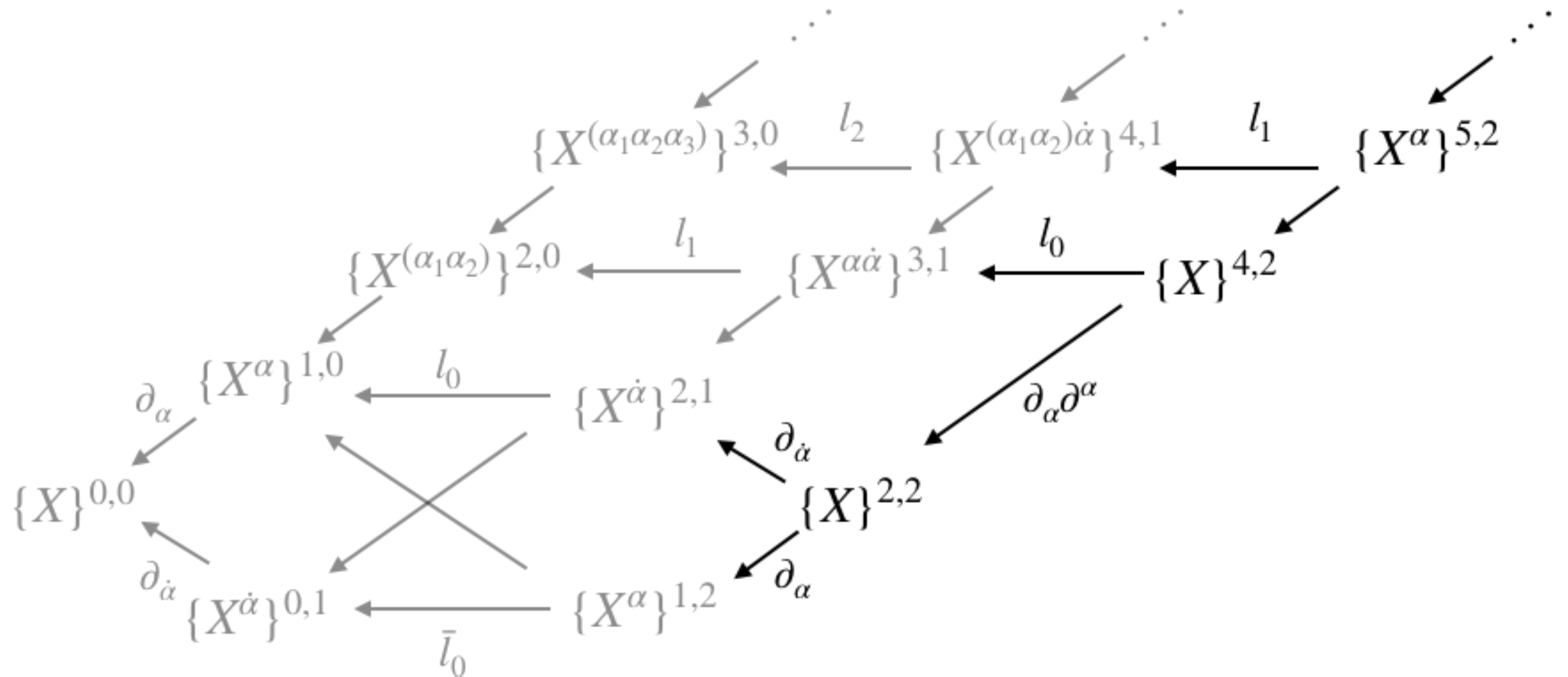}
\end{center}
\caption{$\{X^{(\alpha_1\alpha_2\cdot\cdot\cdot\alpha_{n-4})}\}^{n,2}\ (n\geq2)$ represents the third branch and each term corrects two spaces as can be read from the diagram. The maps $l_n$ are given in the text.}\label{tree3}
\end{figure}

Before summarizing the results, let's pause for the moment and answer the following question: are there any other correction spaces in additional to the three branches (plus the other three accounting for the $\alpha\rightarrow\dot{\alpha}$)? We will prove by contradiction that these six branches are the only corrections that meet the criteria. 

\begin{figure}
\begin{center}
\includegraphics[scale=0.35]{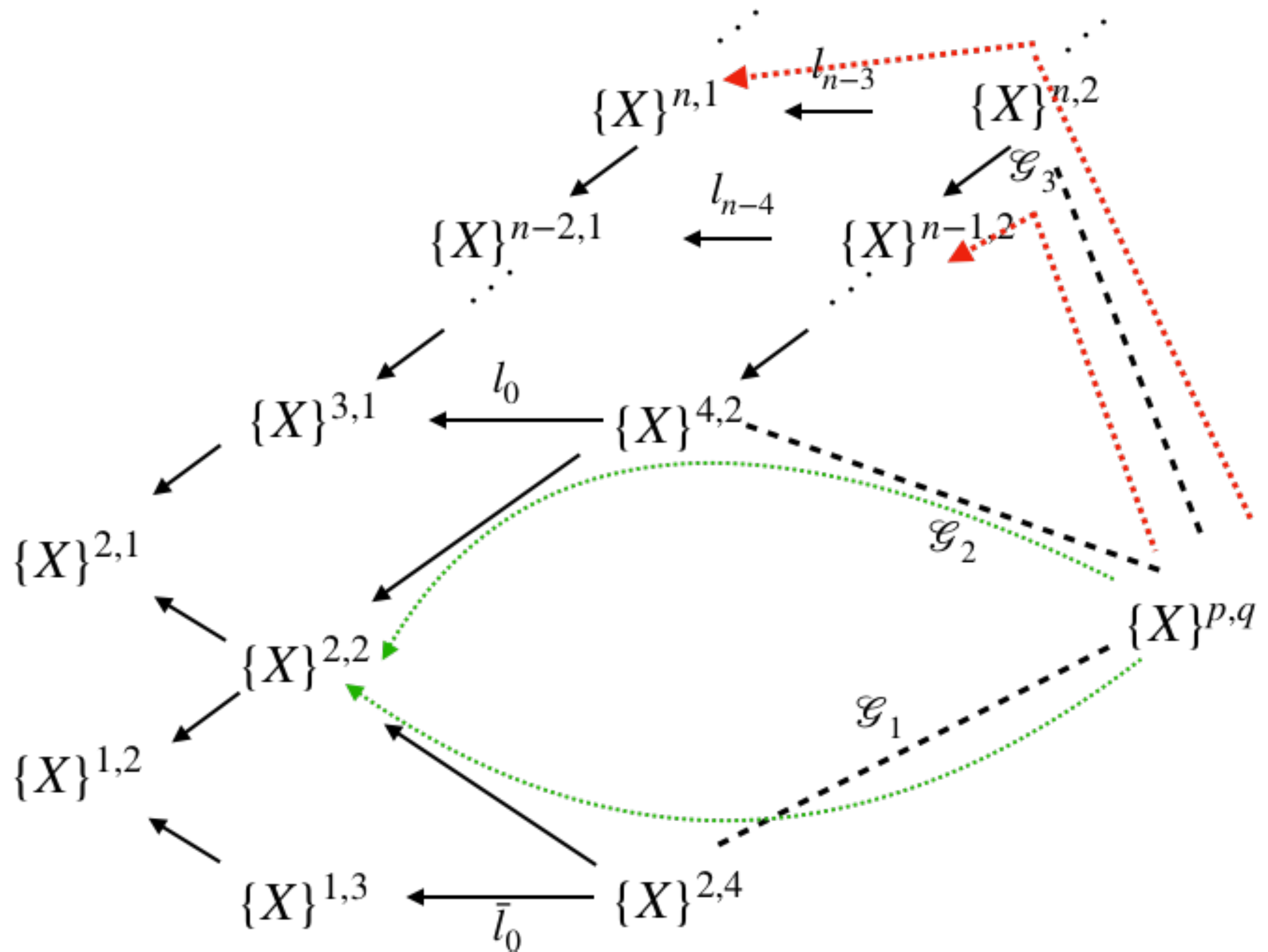}
\end{center}
\caption{If there exist maps $\mathcal{G}_i$, then either green loops or the red lines should be satisfied such that $\{X\}^{p,q}$ meets the criteria of being a correction space.}\label{proof}
\end{figure}

Suppose there exists another correction branch, farther to the right in Fig. \ref{tree}. This branch must begin with some initial space, just as the first branch began with $\{X\}^{1,0}$, the second with $\{X\}^{2,1}$, and the third with $\{X\}^{2,2}$. Initial spaces can be identified because all of their maps connect to lower branches (e.g. $\{X\}^{2,1}$ connects to $\{X\}^{1,0}$ and $\{X\}^{0,1}$). Let us call this initial space ${\{X\}^{p,q}}$. To be a correction space,${\{X\}^{p,q}}$ must satisfy the criteria of Eq.~\eqref{eq:nthorder}; namely it must have a nontrivial map to one of the spaces on the third branch, but all two-step maps must give zero. The possible two-step maps differ depending on whether or not $p = q$.

If $p = q$, the only spaces ${\{X\}^{p,p}}$ can initially map to are $\{X\}^{2,4}$ and $\{X\}^{4,2}$, so, calling the maps $\mathcal G_1, \mathcal G_2$, we have $\mathcal G_1 {\{X\}^{p,p}} \to {\{X\}^{4,2}}$,  $\mathcal G_2 {\{X\}^{p,p}} \to {\{X\}^{2,4}}$. To understand why these are the only possibilities, recall that the maps are combinations of products of $\partial_\alpha, \partial_{\dot\alpha}$, and the composition of the initial space $(\alpha, \dot\alpha)$ indices with the map $(\alpha, \dot\alpha)$ indices must match the target space, so $\mathcal G^{a,b}_1 {\{X\}^{p,p}} \to \{X \}^{p+a, p+b}$. For any space except $\{X\}^{2,4}$ and $\{X\}^{4,2}$, this composition will lead to $|a - b| \ge 3$, a difference between the number of $\partial_\alpha$ and $\partial_{\dot\alpha}$ in $\mathcal G$ of three or more. However, $\partial^3_{\alpha} = \partial^3_{\dot\alpha} = 0$ by Fermi statistics, so any map with $|a - b| \ge 3$ is zero.

Restricted to  $\{X\}^{2,4}$ and $\{X\}^{4,2}$, the possible compound maps we can form are: $\l_0\{X\}^{2,4} = l_0\, \mathcal G_1 \{X\}^{p,p}$, it's complex conjugate (swapping $\mathcal G_1 \leftrightarrow \mathcal G_2$, $l_0 \leftrightarrow \bar{l}_0$), and the loop running through $\{X\}^{2,2}$, $(\partial^2_\alpha \{X\}^{4,2} + \partial^2_{\dot\alpha}\{X\}^{2,4}) = (\partial^2_\alpha \mathcal G_1 + \partial^2_{\dot\alpha} \mathcal G_2)\{X\}^{p,p}$. All three compound maps must vanish if $\{X\}^{p,p}$ is a correction space. To see that this cannot occur, apply $[\{ \partial_{\dot\alpha}, \partial_\alpha \},\partial^\alpha]$ to $\mathcal G_1 \{X\}^{p,p}$. This vanishes, by Eq.~\eqref{eq:useful}. Rewriting  $[\{ \partial_{\dot\alpha}, \partial_\alpha \},\partial^\alpha]\,\mathcal G_1 \{X\}^{p,p}$ using Eq.~\eqref{second branch proof} makes this $(\partial_\alpha l_0 + \partial_{\dot\alpha}\partial^2_{\alpha})\mathcal G_1 \{X\}^{p,p} = 0$. Now, if we take $l_0\, \mathcal G_1 \{X\}^{p,p} = 0$ to satisfy the correction criteria, then $\partial_{\dot\alpha}\partial^2_{\alpha}\mathcal G_1 \{X\}^{p,p} = 0$. As a result, we can conclude $\partial^2_{\alpha}\mathcal G_1 \{X\}^{p,p}$ must be chiral, as it is annihilated by $\partial_{\dot\alpha}$ -- but it must also be antichiral as $\partial^3_{\alpha}\mathcal G_1 \{X\}^{p,p} = 0$. The analogous logic holds for $\mathcal G_2\,\{X\}^{p,p}$, swapping $\bar{l}_0 \leftrightarrow l_0$.  As only a constant field be both chiral and antichiral, $\{X\}^{p,p}$ cannot be a correction space.

If $p \ne q$, ${\{X\}^{p,q}}$ can initially map onto spaces higher up on the third branch. Assuming without loss of generality that this space is on the `$\partial_\alpha$' side, and calling the initial map $\mathcal G_3$, we have $\mathcal G_3 {\{X\}^{p,q}} \to {\{X\}^{n,2}}$ for $n \ge 4$. From $\{X\}^{n,2}$ there are two compound maps we can form, a horizontal map to $\{X\}^{n-1,1}$ via $l_{n-3}$, or a map in the $\partial_\alpha$ direction taking us to $\{X\}^{n-1,2}$. For both to vanish we need:
\begin{align}
& \partial_\alpha \{X\}^{n,2} = \partial_\alpha\, \mathcal G_3 \{X\}^{p,q} = 0 \nonumber \\
& l_{n-3} \{X\}^{n,2} = l_{n-3} \mathcal G_3 \{X\}^{p,q} = 0 \nonumber 
\end{align}
The first of these conditions requires $\mathcal G_3 \{X\}^{p,q} $ to be antichiral. The second requires $\mathcal G_3 \{X\}^{p,q}$ to be chiral, which we can see by expanding $l_{n-3} = (a\, \partial_{\alpha\dot{\alpha}} + b\, \partial_{\dot{\alpha}\alpha})$ for constants $a,b$ (see Eq.~\eqref{eq:aandb}), and applying it to an antichiral $\mathcal G_3 \{X\}^{p,q}$. We are left with $\partial_{\alpha\dot{\alpha}} \mathcal G_3 \{X\}^{p,q} = 0$  -- implying $\mathcal G_3 \{X\}^{p,q}$ must also be chiral.  Thus, $\{X\}^{p,q}$, $p \ne q$ also cannot satisfy the correction requirements and we conclude that the three branches (plus their complex conjugates) are the only possible correction spaces.

\subsubsection{Summing the corrections}\label{summing}

Changing all dotted indices to undotted (and vice versa) on the three correction branches studied above gets us the three `$\partial_{\dot\alpha}$' branches, making the full set of corrections six branches.  To summarize, the corrections are given by:\\

\noindent (i) \ \ $0^{\text {th}}$ order corrections: $\{X\}^{1,0},\ \{X\}^{0,1}$\\
(ii) \ $1^{\text {st}}$ order corrections: $\{X\}^{1,2},\ \{X\}^{2,1},\ \{X\}^{0,2},\ \{X\}^{2,0}$\\
(iii) $2^{\text {nd}}$ order corrections: $\{X\}^{1,3},\ \{X\}^{3,1},\ \{X\}^{0,3},\ \{X\}^{3,0},\ \{X\}^{2,2}$\\
(iv) $(n\geq3)^{\text {th}}$ order corrections: $\{X\}^{1,n+1},\ \{X\}^{n+1,1},\ \{X\}^{0,n+1},\ \{X\}^{n+1,0},\ \{X\}^{n+1,2},\ \{X\}^{2,n+1}$.\\

\noindent While the correction branches theoretically extend `to infinity', in practice they are limited by the number of derivatives in the original $\mathcal O$ operator. For example, consider an operator with two $\partial_\alpha$ and two $\partial_{\dot\alpha}$. As each correction space involves terms with one fewer derivative, this operator can only have zeroth and first order corrections. So, while the corrections terminate at four derivatives for an operator in a non-supersymmetric theory regardless of how many derivatives that operator has, the correction order to an operator in a supersymmetric theory match the derivative order, e.g. an $\mathcal O(\partial^n_\alpha\partial^n_{\dot\alpha})$ operator will have $n^{\text {th}}$ order corrections.
Fermi statistics can also come into play when determining which spaces are populated, as we will see in the second example in next section.

Combining these corrections into Eq.~\eqref{num}, the number of independent operators including all IBP corrections is given by:
\begin{align}\label{master0}
\# \ of \ independent\  operators &=\# \{{X}\}^{0,0}\\
&-\# (\{{X}\}^{0,1}+\{{X}\}^{1,0})\nonumber\\
&+\# (\{X\}^{1,2}+\{{X}\}^{2,1}+\{{X}\}^{0,2}+\{{X}\}^{2,0})\nonumber\\
&-\# (\{X\}^{1,3}+\{{X}\}^{3,1}+\{{X}\}^{0,3}+\{{X}\}^{3,0}+\{{X}\}^{2,2})\nonumber\\
&+\# (\{X\}^{1,4}+\{{X}\}^{4,1}+\{{X}\}^{0,4}+\{{X}\}^{4,0}+\{{X}\}^{4,2}+\{{X}\}^{2,4})\nonumber \\
&\cdots \nonumber
\end{align}
Using the character formulae of the spaces $\{X\}^{p,q}$ summarized in Table.~\ref{table} and making use of \eqref{num}, the total dressed prefactor is \footnote{The function $f^{(p,q)}$ is defined such that even order correction spaces, i.e. $S^{2k}_i$, carry minus signs, as we see in \eqref{num}. Table \ref{table} provides the translation rule from $S^n_i$ to $\{X^{p,q}\}$.}
\begin{equation}\label{dressed}
\begin{split}
&1-\sum_{n=0}^{\infty}(-1)^n \{S_i^n\}\chi_{S_i^n}^{}\\
=&\sum (-1)^{f(p,q)} P^pQ^q\chi_{X^{p,q}}^{}\\=&1\\
-&(Px+Qy)\\
+&(PQ^2x+P^2Qy+P^2(x^2-1)+Q^2(y^2-1))\\
-&(PQ^3xy+P^3Qxy+P^3(x^3-2x)+Q^3(y^3-2y)+P^2Q^2)\\
+&\cdots
\end{split}
\end{equation}
Formally we can calculate the sum of this infinite series, and the results are given in \eqref{formalsum1}. When R-symmetry is considered, one should replace $P\rightarrow P'=Pz^{-1}$ and $Q\rightarrow Q'=Qz$.

We want to encode this into the Hilbert series, which then automatically projects out the number of operators in each space. The easiest way to do it is by adding the corresponding character of each space into the integrand \eqref{general hilbert series}. Putting everything together, our master formula is:
\begin{equation}\label{master}
\mathcal{H}(P,Q,\{\Phi_i\})=\int d\mu_{Lorentz}d\mu_{gauge}d\mu_{U(1)_R}[\sum (-1)^{f(p,q)} (Pz^{-1})^p(Qz)^q\chi_{X^{p,q}}^{}]PE[\sum_i \Phi_i\tilde{\chi}_{R,i}^{}], 
\end{equation}
where $\tilde{\chi}_{R,i}^{}$ contains Lorentz, gauge and R-symmetry characters. One should keep in mind that we are actually working with superfields, not non-supersymmetric fields in this paper. Therefore the $\sum\limits_{i=1}^N\Phi_i\tilde{\chi}_{R,i}^{}$ should be understood as the sum over both lowest component fields and the next-order terms with one more derivative, which represent the bosonic (fermionic) superpartners of the lowest component fields. For example, for a chiral superfield $\Phi$, the argument in the PE is given by the sum $\Phi\tilde{\chi}_{(0,0)}+P(D\Phi)\tilde{\chi}_{(\frac{1}{2},0)}$. This is a manifestation of supersymmetry in the sense that one bosonic degree of freedom is related to one fermionic degree of freedom, and vice versa.

\subsection{Examples}\label{example}
In this section, we will first go through an example to demonstrate the general procedure to count the operators, then we'll look at two other examples with non-vanishing higher representations. 

Consider the single flavor case with two chiral superfields and two anti-chiral superfields, i.e. $\Phi^2\Phi^{\dagger 2}$. We want to find the operator basis for operator spaces in the form of $\partial_\alpha^n\partial_{\dot{\alpha}}^n\Phi^2\Phi^{\dagger 2}$, where $n$ is the number of superderivatives $\partial_\alpha$ and $\partial_{\dot{\alpha}}$. The first non-trivial space generated by operator in this form is the $\partial_\alpha^2\partial_{\dot{\alpha}}^2\Phi^2\Phi^{\dagger 2}$ at dimension 6 when $n=2$. It's not difficult to see that there are 6 operators in the space modulo EOM, namely:
\begin{align}
(\partial_\alpha\Phi)(\partial^\alpha\Phi) (\partial_{\dot{\alpha}}\Phi^\dagger)(\partial^{\dot{\alpha}}\Phi^\dagger) \quad & \quad (\partial_{\dot{\alpha}}\partial_\alpha\Phi)(\partial^{\dot{\alpha}}\partial^\alpha\Phi)(\Phi^\dagger)^2 \nonumber \\
\Phi^2(\partial_\alpha\partial_{\dot{\alpha}}\Phi^\dagger)(\partial^\alpha\partial^{\dot{\alpha}}\Phi^\dagger)\quad & \quad \Phi\Phi^\dagger(\partial_{\dot{\alpha}}\partial_\alpha\Phi)(\partial^\alpha\partial^{\dot{\alpha}}\Phi^\dagger)\nonumber \\ \Phi^\dagger(\partial^\alpha\Phi)(\partial_{\dot{\alpha}}\partial_\alpha\Phi)(\partial^{\dot{\alpha}}\Phi^\dagger)\quad & \quad \Phi(\partial^{\dot{\alpha}}\Phi^\dagger)(\partial_\alpha\partial_{\dot{\alpha}}\Phi^\dagger)(\partial^{\dot{\alpha}}\Phi^\dagger).
\end{align}

However, not all of them are independent under IBP relations. For example, the equation
\begin{equation}
\partial_\alpha[\Phi(\partial^\alpha\Phi)(\partial_{\dot{\alpha}}\Phi^\dagger)(\partial^{\dot{\alpha}}\Phi^\dagger)]\sim (\partial_\alpha\Phi)(\partial^\alpha\Phi) (\partial_{\dot{\alpha}}\Phi^\dagger)(\partial^{\dot{\alpha}}\Phi^\dagger)+2\Phi(\partial_\alpha\Phi)(\partial^\alpha\partial^{\dot{\alpha}}\Phi^\dagger)(\partial_{\dot{\alpha}}\Phi^\dagger)
\end{equation}
indicates that the two terms on the right hand side are not independent, and as a result only one of them should exist in the operator space. Notice that terms like $\Phi(\partial_\alpha^2\Phi)(\partial_{\dot{\alpha}}\Phi^\dagger)^2$ are dropped due to the EOM relations in the above equation. Since there are $(3+3)$ terms in $\{X\}^{1,0}$ and $\{X\}^{1,0}$ spaces, we can derive six IBP relations:
\begin{equation}
\begin{split}
&\partial_\alpha[\Phi(\partial^\alpha\Phi)(\partial_{\dot{\alpha}}\Phi^\dagger)(\partial^{\dot{\alpha}}\Phi^\dagger)]\sim (\partial_\alpha\Phi)(\partial^\alpha\Phi) (\partial_{\dot{\alpha}}\Phi^\dagger)(\partial^{\dot{\alpha}}\Phi^\dagger)-2\Phi(\partial_\alpha\Phi)(\partial^\alpha\partial^{\dot{\alpha}}\Phi^\dagger)(\partial_{\dot{\alpha}}\Phi^\dagger),\\
&\partial_\alpha[\Phi^2(\partial_{\dot{\alpha}}\Phi^\dagger)(\partial^\alpha\partial^{\dot{\alpha}}\Phi^\dagger)]\sim 2\Phi(\partial_\alpha\Phi)(\partial^\alpha\partial^{\dot{\alpha}}\Phi^\dagger)(\partial_{\dot{\alpha}}\Phi^\dagger)+\Phi^2(\partial_\alpha\partial_{\dot{\alpha}}\Phi^\dagger)(\partial^\alpha\partial^{\dot{\alpha}}\Phi^\dagger),\\
&\partial_\alpha[\Phi(\partial^{\dot{\alpha}}\partial^\alpha\Phi)\Phi^\dagger(\partial_{\dot{\alpha}}\Phi^\dagger)]\sim (\partial^{\dot{\alpha}}\partial^\alpha\Phi)(\partial_\alpha\Phi)\Phi^\dagger(\partial_{\dot{\alpha}}\Phi^\dagger)+
\Phi(\partial^{\dot{\alpha}}\partial^\alpha\Phi)\Phi^\dagger(\partial_\alpha\partial_{\dot{\alpha}}\Phi^\dagger),\\
&\partial_{\dot{\alpha}}[(\partial_\alpha\Phi)(\partial^\alpha\Phi)\Phi^\dagger(\partial^{\dot{\alpha}}\Phi^\dagger)]\sim (\partial_\alpha\Phi)(\partial^\alpha\Phi) (\partial_{\dot{\alpha}}\Phi^\dagger)(\partial^{\dot{\alpha}}\Phi^\dagger)+2(\partial_{\dot{\alpha}}\partial_\alpha\Phi)(\partial^\alpha\Phi)\Phi^\dagger(\partial^{\dot{\alpha}}\Phi^\dagger),\\
&\partial_{\dot{\alpha}}[(\partial_\alpha\Phi)(\partial^{\dot{\alpha}}\partial^\alpha\Phi)\Phi^{\dagger2}]\sim(\partial_{\dot{\alpha}}\partial_\alpha\Phi)(\partial^{\dot{\alpha}}\partial^\alpha\Phi)\Phi^{\dagger2}-2(\partial_\alpha\Phi)(\partial^{\dot{\alpha}}\partial^\alpha\Phi)\Phi^{\dagger}(\partial_{\dot{\alpha}}\Phi^\dagger),\\
&\partial_{\dot{\alpha}}[\Phi(\partial_\alpha\Phi)\Phi^\dagger(\partial^\alpha\partial^{\dot{\alpha}}\Phi^\dagger)]\sim \Phi(\partial_{\dot{\alpha}}\partial_\alpha\Phi)\Phi^\dagger(\partial^\alpha\partial^{\dot{\alpha}}\Phi^\dagger)-\Phi(\partial_\alpha\Phi)(\partial^\alpha\partial^{\dot{\alpha}}\Phi^\dagger)(\partial_{\dot{\alpha}}\Phi^\dagger).
\end{split}
\end{equation}

A brute force way to find the number of independent operators is to apply matrix rank approach, as in Ref.~\cite{Lehman:2015coa}. Specifically, we label each different element as $x_i$ and each relation provides an equation $\sum_j a_jx_j=0$, where $x_j$ are the elements in each relation. For example the previous six relations can be transformed into a set of equations: $x_1-2x_2=0,\ 2x_2+x_3=0,\ x_4+x_5=0,\ x_1+2x_4=0,\ x_6-2x_4=0,\ x_5-x_2=0$. What's left is to find the solution to this set of equations and the easiest way is to use a matrix solution. The rank of the matrix then gives the independent number of $x_i$'s. Rather than directly calculate the rank of the matrix relations for this example, we will proceed more methodically and check how the existence of previous (meaning fewer derivatives, so to the right in Fig.~\ref{tree}) spaces imply dependancies among the relations at each step along the way.

Applied to the case at hand, it seems like that the six relations will remove all six terms in the $\{X\}^{0,0}$. However, this is not true because the six relations are not independent, and therefore we need higher order corrections coming from $\{X^{\alpha}\}^{1,2}$ and $\{X^{\dot{\alpha}}\}^{2,1}$. The relations we get from these two spaces are:
\begin{align}
&\partial_\alpha[\Phi(\partial_\alpha\Phi)(\partial_{\dot{\alpha}}\Phi^\dagger)^2]+2\partial_\alpha[\Phi(\partial_{\dot{\alpha}}\partial_\alpha\Phi)\Phi^\dagger(\partial_{\dot{\alpha}}\Phi^\dagger)]\sim\partial_{\dot{\alpha}}[(\partial_\alpha\Phi)^2\Phi^\dagger(\partial_{\dot{\alpha}}\Phi^\dagger)]+2\partial_{\dot{\alpha}}[(\partial_\alpha\Phi)^2\Phi^\dagger(\partial_{\dot{\alpha}}\Phi^\dagger)] \nonumber \\
&\partial_{\dot{\alpha}}[(\partial_\alpha\Phi)^2\Phi^\dagger(\partial_{\dot{\alpha}}\Phi^\dagger)]+2\partial_{\dot{\alpha}}[(\partial_\alpha\Phi)^2\Phi^\dagger(\partial_{\dot{\alpha}}\Phi^\dagger)]\sim \partial_\alpha[\Phi(\partial_\alpha\Phi)(\partial_{\dot{\alpha}}\Phi^\dagger)^2]+2\partial_\alpha[\Phi(\partial_{\dot{\alpha}}\partial_\alpha\Phi)\Phi^\dagger(\partial_{\dot{\alpha}}\Phi^\dagger)] \nonumber 
\end{align}
Clearly, the above relations are the same, so we should add 1 back instead of 2. The reason comes from the fact that the LHS of the above two equations comes from the term in $\{X\}^{2,2}$, and therefore are corrected by $\{X\}^{2,2}$:
\begin{equation}
\begin{split}
&l_0\partial_\alpha[\Phi^2\Phi^{\dagger2}]\sim (\partial_\alpha)^2\partial_{\dot{\alpha}}[\Phi^2\Phi^{\dagger2}],\\
&\overline{l}_0\partial_{\dot{\alpha}}[\Phi^2\Phi^{\dagger2}]\sim (\partial_{\dot{\alpha}})^2\partial_\alpha[\Phi^2\Phi^{\dagger2}].
\end{split}
\end{equation}

\begin{figure}
\begin{subfigure}{.45\textwidth}
  \centering
  \includegraphics[width=1\linewidth]{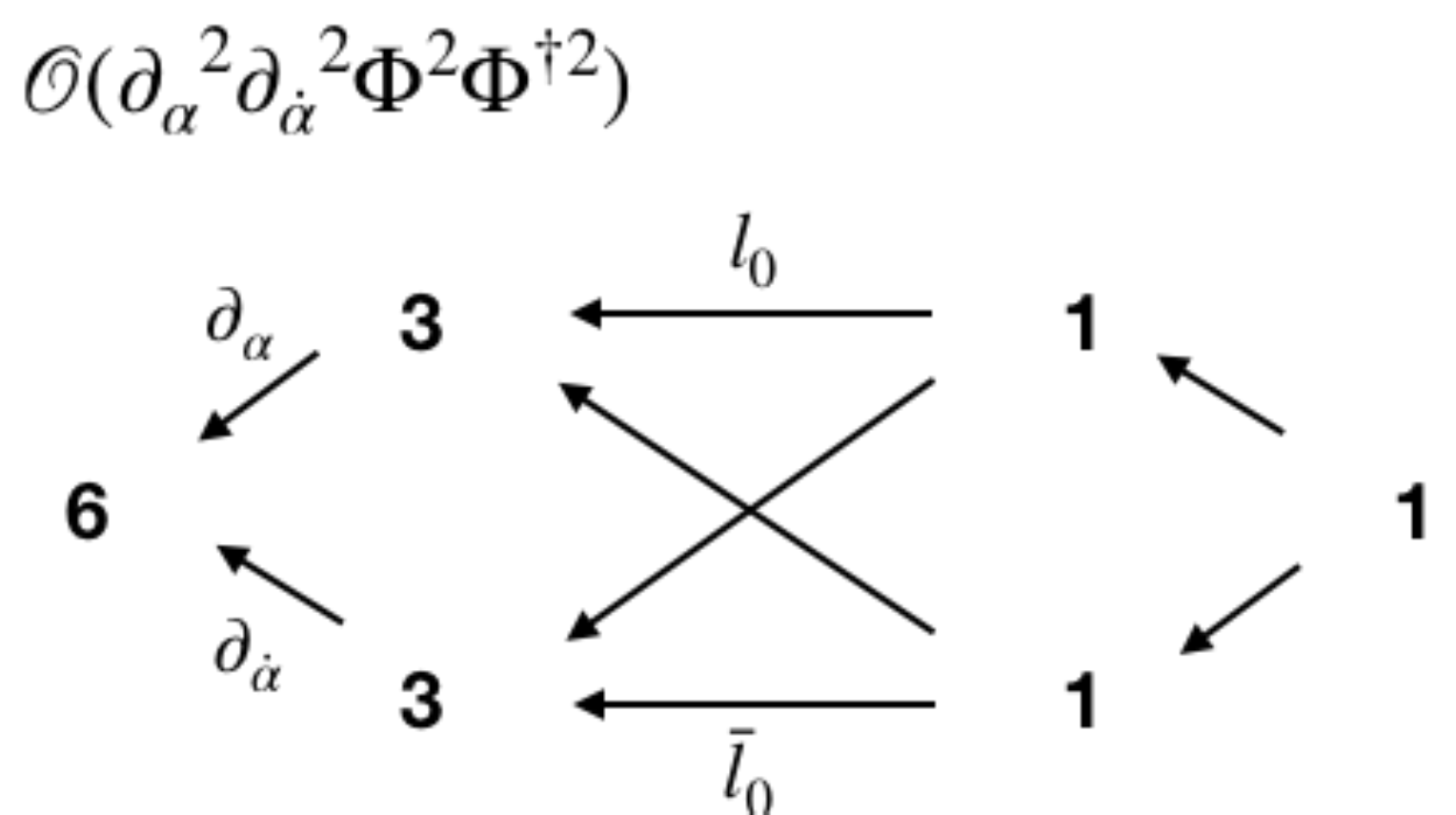}  
\end{subfigure}
\begin{subfigure}{.5\textwidth}
  \centering
  \includegraphics[width=1\linewidth]{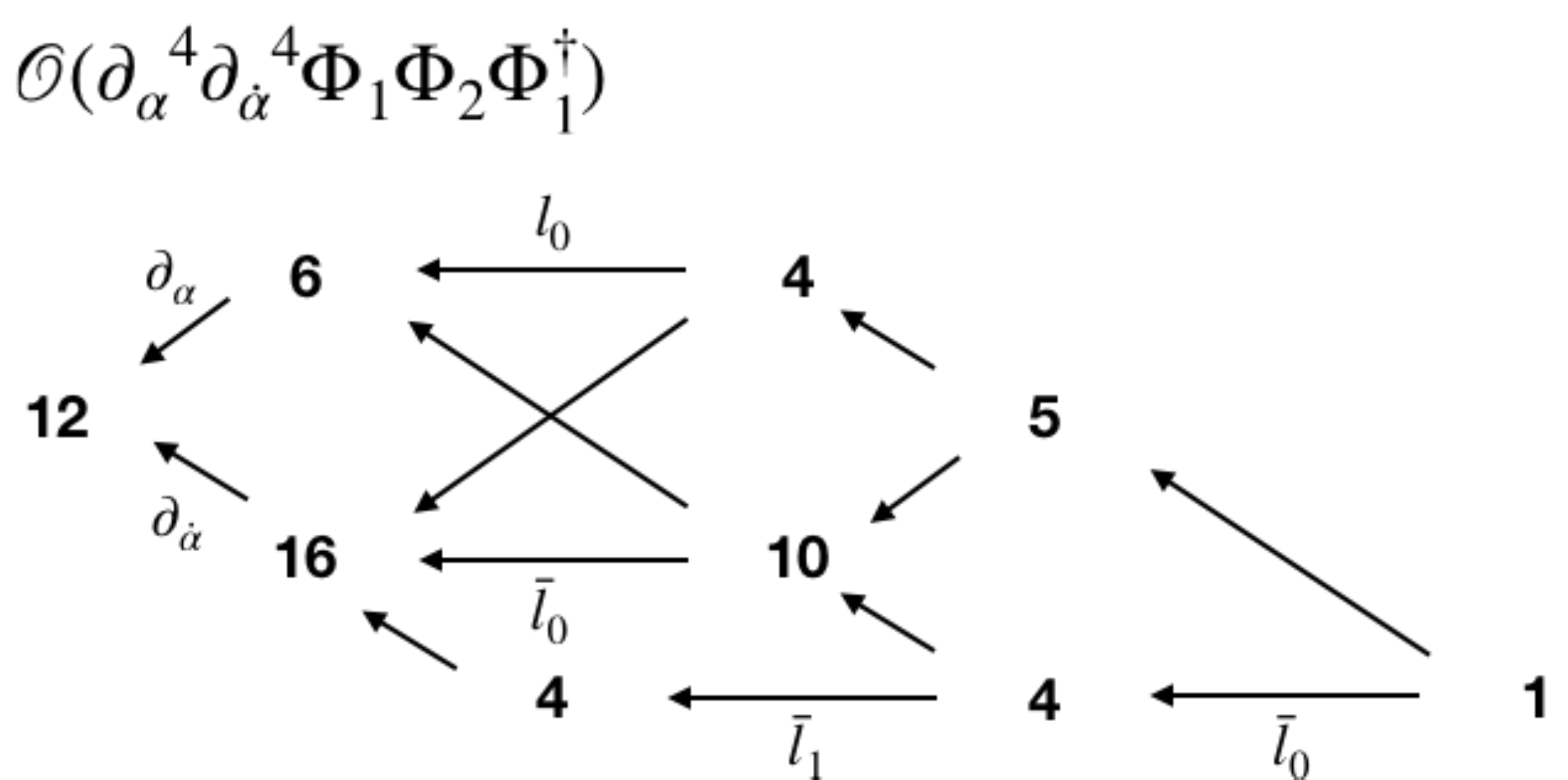}  
\end{subfigure}
\caption{These two tree diagrams represent the number in each space of $\partial_\alpha^2\partial_{\dot{\alpha}}^2\Phi^2\Phi^{\dagger 2}$ and $\mathcal{O}(\partial_\alpha^4\partial_{\dot{\alpha}}^4\Phi_1\Phi_2\Phi_1^\dagger)$. Null spaces are not shown here.}
\label{example12}
\end{figure}

Finally we add up the numbers: $6-3-3+1+1-1=1$, which gives the correct number of independent operators. The tree diagram is given on the left in Fig \ref{example12}.

Since there are only four superderivatives in this example, spaces transforming under larger representations are automatically null. However, once we consider operators with more superderivatives, those spaces must be taken into account. Non-vanishing larger representations are very common especially when we deal with multiple flavors, since in that case the constraints from Fermi statistics is weaker.

To illustrate this point, let's take a look at the second example, involving two flavors and eight superderivatives, $\mathcal{O}(\partial_\alpha^4\partial_{\dot{\alpha}}^4\Phi_1\Phi_2\Phi_1^\dagger)$. We will not go into details to write down all orders of corrections, instead, we will compare the results we get from brute force way and our approach.

The tree diagram is shown on the right in Fig.~\ref{example12}. From IBP relations we get the following equations:
\begin{equation}
\begin{split}
&\partial_{\dot{\alpha}}(\partial_{\alpha\dot{\beta}\beta}\Phi_1\partial^{\dot{\alpha}\alpha\dot{\beta}\beta}\Phi_2\Phi_1^\dagger)\ \ \sim\partial_{\dot{\alpha}\alpha\dot{\beta}\beta}\Phi_1\partial^{\dot{\alpha}\alpha\dot{\beta}\beta}\Phi_2\Phi_1^\dagger-\partial_{\alpha\dot{\beta}\beta}\Phi_1\partial^{\dot{\alpha}\alpha\dot{\beta}\beta}\Phi_2\partial_{\dot{\alpha}}\Phi_1^\dagger\\
&\partial_{\dot{\alpha}}(\partial_{\alpha\dot{\beta}\beta}\Phi_1\partial^{\alpha\dot{\beta}\beta}\Phi_2\partial^{\dot{\alpha}}\Phi_1^\dagger)\sim\partial_{\dot{\alpha}\alpha\dot{\beta}\beta}\Phi_1\partial^{\alpha\dot{\beta}\beta}\Phi_2\partial^{\dot{\alpha}}\Phi_1^\dagger+\partial_{\alpha\dot{\beta}\beta}\Phi_1\partial^{\dot{\alpha}\alpha\dot{\beta}\beta}\Phi_2\partial_{\dot{\alpha}}\Phi_1^\dagger\\
&\partial_{\dot{\alpha}}(\partial_{\alpha\dot{\beta}\beta}\Phi_1\partial^{\dot{\beta}\beta}\Phi_2\partial^{\alpha\dot{\alpha}}\Phi_1^\dagger)\sim\partial_{\dot{\alpha}\alpha\dot{\beta}\beta}\Phi_1\partial^{\dot{\beta}\beta}\Phi_2\partial^{\alpha\dot{\alpha}}\Phi_1^\dagger\\
&\partial_{\dot{\alpha}}(\partial_{\alpha\dot{\beta}\beta}\Phi_1\partial^{\dot{\alpha}\alpha}\Phi_2\partial^{\beta\dot{\beta}}\Phi_1^\dagger)\sim\partial_{\dot{\alpha}\alpha\dot{\beta}\beta}\Phi_1\partial^{\dot{\alpha}\alpha}\Phi_2\partial^{\beta\dot{\beta}}\Phi_1^\dagger-\partial_{\alpha\dot{\beta}\beta}\Phi_1\partial^{\dot{\alpha}\alpha}\Phi_2\partial_{\dot{\alpha}}^{\ \beta\dot{\beta}}\Phi_1^\dagger\\
&\partial_{\dot{\alpha}}(\partial_{\alpha\dot{\beta}\beta}\Phi_1\partial^{\alpha}\Phi_2\partial^{\dot{\alpha}\beta\dot{\beta}}\Phi_1^\dagger)\sim\partial_{\dot{\alpha}\alpha\dot{\beta}\beta}\Phi_1\partial^{\alpha}\Phi_2\partial^{\dot{\alpha}\beta\dot{\beta}}\Phi_1^\dagger-\partial_{\alpha\dot{\beta}\beta}\Phi_1\partial_{\dot{\alpha}}^{\ \alpha}\Phi_2\partial^{\dot{\alpha}\beta\dot{\beta}}\Phi_1^\dagger\\
&\partial_{\dot{\alpha}}(\partial_{\alpha\dot{\beta}\beta}\Phi_1\Phi_2\partial^{\alpha\dot{\alpha}\beta\dot{\beta}}\Phi_1^\dagger)\ \ \sim\partial_{\dot{\alpha}\alpha\dot{\beta}\beta}\Phi_1\Phi_2\partial^{\alpha\dot{\alpha}\beta\dot{\beta}}\Phi_1^\dagger\\
&\partial_{\dot{\alpha}}(\partial^{\dot{\alpha}\alpha\dot{\beta}\beta}\Phi_1\partial_\alpha\Phi_2\partial_{\beta\dot{\beta}}\Phi_1^\dagger)\sim\partial^{\dot{\alpha}\alpha\dot{\beta}\beta}\Phi_1\partial_{\dot{\alpha}\alpha}\Phi_2\partial_{\beta\dot{\beta}}\Phi_1^\dagger-\partial^{\dot{\alpha}\alpha\dot{\beta}\beta}\Phi_1\partial_\alpha\Phi_2\partial_{\dot{\alpha}\beta\dot{\beta}}\Phi_1^\dagger\\
&\partial_{\dot{\alpha}}(\partial_{\dot{\beta}\beta}\Phi_1\partial_\alpha\Phi_2\partial^{\dot{\alpha}\alpha\dot{\beta}\beta}\Phi_1^\dagger)\sim\partial_{\dot{\beta}\beta}\Phi_1\partial_{\dot{\alpha}\alpha}\Phi_2\partial^{\dot{\alpha}\alpha\dot{\beta}\beta}\Phi_1^\dagger\\
&\partial^\alpha(\partial_{\dot{\alpha}\alpha\dot{\beta}\beta}\Phi_1\partial^{\dot{\alpha}\beta}\Phi_2\partial^{\dot{\beta}}\Phi_1^\dagger)\sim\partial_{\dot{\alpha}\alpha\dot{\beta}\beta}\Phi_1\partial^{\alpha\dot{\alpha}\beta}\Phi_2\partial^{\dot{\beta}}\Phi_1^\dagger+\partial_{\dot{\alpha}\alpha\dot{\beta}\beta}\Phi_1\partial^{\dot{\alpha}\beta}\Phi_2\partial^{\alpha\dot{\beta}}\Phi_1^\dagger\\
&\partial^\alpha(\partial_{\dot{\alpha}\alpha\dot{\beta}\beta}\Phi_1\Phi_2\partial^{\dot{\alpha}\beta\dot{\beta}}\Phi_1^\dagger)\ \ \sim\partial_{\dot{\alpha}\alpha\dot{\beta}\beta}\Phi_1\partial^\alpha\Phi_2\partial^{\dot{\alpha}\beta\dot{\beta}}\Phi_1^\dagger+\partial_{\dot{\alpha}\alpha\dot{\beta}\beta}\Phi_1\Phi_2\partial^{\alpha\dot{\alpha}\beta\dot{\beta}}\Phi_1^\dagger\\
&\partial^\alpha(\partial_{\dot{\alpha}\beta}\Phi_1\partial_{\dot{\beta}\alpha}\Phi_2\partial^{\dot{\alpha}\beta\dot{\beta}}\Phi_1^\dagger)\sim\partial^\alpha_{\ \dot{\alpha}\beta}\Phi_1\partial_{\dot{\beta}\alpha}\Phi_2\partial^{\dot{\alpha}\beta\dot{\beta}}\Phi_1^\dagger+\partial_{\dot{\alpha}\beta}\Phi_1\partial_{\dot{\beta}\alpha}\Phi_2\partial^{\alpha\dot{\alpha}\beta\dot{\beta}}\Phi_1^\dagger
\end{split}
\end{equation}
where $\partial_{abc\cdot\cdot\cdot}$ is understood as $\partial_a\partial_b\partial_c\cdot\cdot\cdot$, $a,b,c\in\{\alpha,\dot{\alpha},\beta,\dot{\beta}\}$. We only list half of the 22 relations and the other 11 can be found by exchanging $\Phi_1\leftrightarrow\Phi_2$. Once one writes down all 22 relations, it becomes easy to calculate the rank of the giant matrix to find the number of independent terms. Plugged this into any math software, we find 0, exactly the same result you get by applying \eqref{master}: $12-16-6+4+10+4-5-4+1=0$.

\begin{figure}
\begin{center}
\includegraphics[scale=0.35]{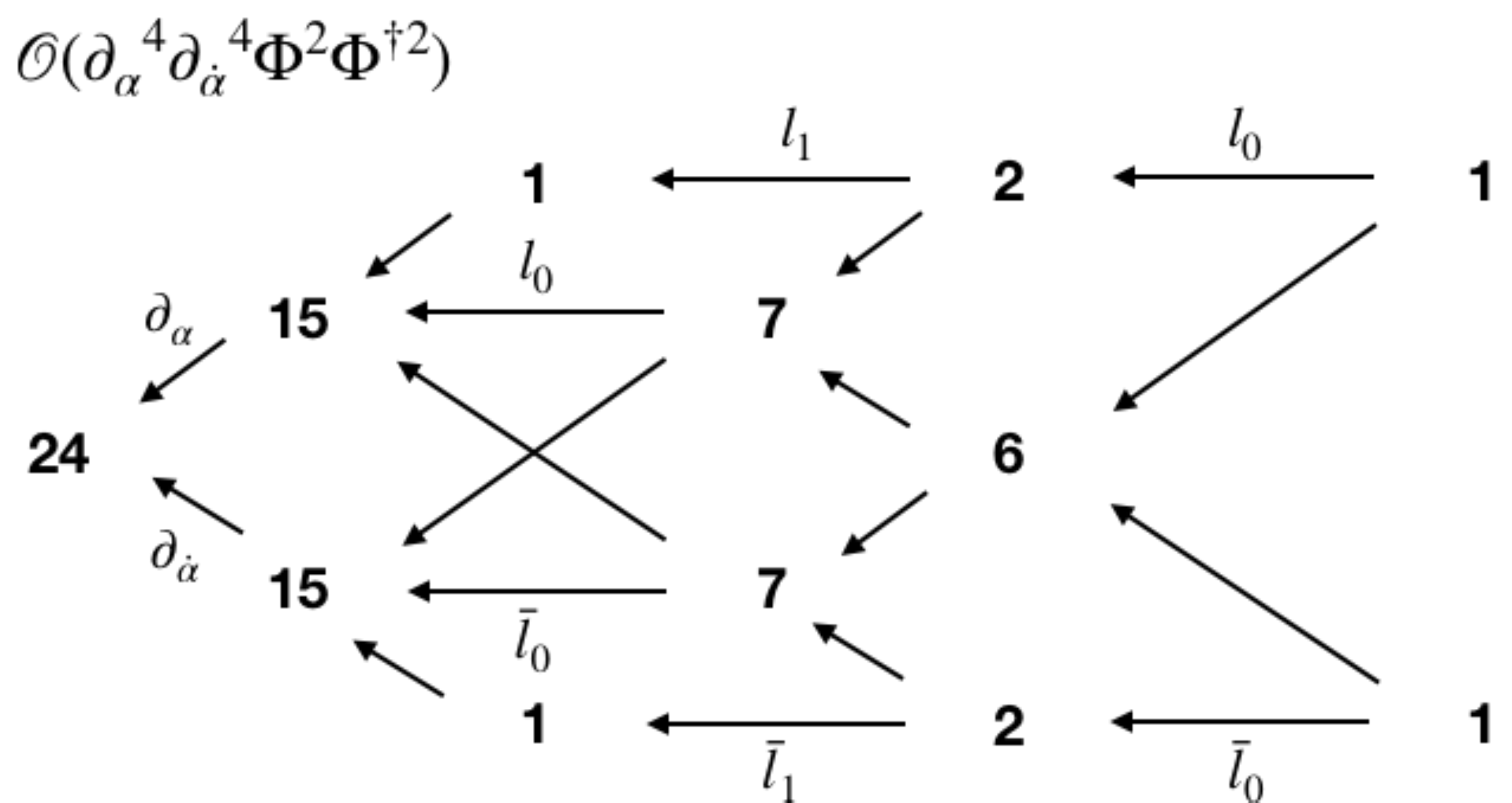}
\end{center}
\caption{This is the tree diagram for $\partial_\alpha^4\partial_{\dot{\alpha}}^4\Phi^2\Phi^{\dagger 2}$, all null spaces are removed.}\label{example3}
\end{figure}

A third example is given in Fig.~\ref{example3}, where in this case we have four superfields and eight superderivatives. It is expected that as the dimension increases, the number of non-vanishing spaces increases, and the related tree diagram extends further. In this example, you can see that if one tries to find the number of independent operators, the dimension of the matrix is $30\times24$. As we add more and more superfields and superderivatives, the matrix becomes larger and is almost impossible for one to solve by brute force. Our approach provides a more accessible way since you no longer need to really find the matrix, instead you calculate the number of possible operators in each space, which can be easily done using Hilbert series. In this case we find $24-15-15+1+1+7+7-2-6-2+1+1=2$ invariants.

\section{Conclusion and discussion}\label{Conclusion}

In this paper we have extended the technique of counting the number of independent effective operators  to a $N=1$ supersymmetric theory with chiral superfields. Hilbert series and the plethystic exponential provide a way to form the full operator spaces, while EOM and IBP relations are removed by manipulating the characters and considering the correction spaces. 

We find that supersymmetry, and in particular the presence of two, anticommuting derivatives, lead to interesting differences with respect to the non-supersymmetric case. The two superderivatives give two initial (`zeroth order') IBP relations, and the structure of the higher order IBP corrections becomes more complicated. We find the corrections organize themselves into six branches, grouped into one set of three oriented in the $\partial_\alpha$ direction and a complex conjugate set oriented in the $\partial_{\dot\alpha}$ direction. All correction spaces satisfy a master relation, Eq.~\eqref{eq:nthorder}. 

While a bit formal at first glance, we have seen that the interrelations among the maps implied by Eq.~\eqref{eq:nthorder} are connected to relations between the two superderivatives $\partial_\alpha, \partial_{\dot\alpha}$ and the `normal' (or `bosonic') derivative $\partial_\mu$. As the superderivatives are fermionic, corrections spaces with multiple $SU(2)_L$ or $SU(2)_R$ indices are required to be symmetric in those indices. As there is no barrier to forming symmetric tensors of arbitrarily high dimension, the correction branches formally extend infinitely, though in practice they are cut by the number of derivatives in the operator we care about correcting. This can be contrasted with correction spaces in non-supersymmetric spaces, which must be antisymmetric in Lorentz indices and this truncate with $\{X\}^{[\mu\nu\rho\sigma]}$ regardless of how many derivatives the operator we want to correct has.

The full set of corrections can be resummed and turned into a prefactor, namely \eqref{dressed},\ $\sum (-1)^{f(p,q)} P^pQ^q\chi_{X^{p,q}}^{}$, to insert into the Haar measure integral along with the plethystic exponential of the chiral/antichiral superfields of interest. This prefactor plays the role of the $1/P$ factor~\cite{Henning:2017fpj}  in non-supersymmetric theories, and enacts the subtraction of IBP relations by projecting out and counting the appropriate correction spaces. For non-supersymmetric theories, Ref.~\cite{Henning:2017fpj} showed how the $1/P$ factor could be obtained by considering conformal symmetry as the organizing principle underlying how fields are combined, rather than Lorentz symmetry. The non-orthonormality of the (non-compact) conformal group characters, combined with integration over the dilatation portion of the Haar measure/Cartan subalgebra then yields $1/P$.  In that light, a natural question to ask is whether a  generalization using superconformal representations reproduces $\sum (-1)^{f(p,q)} P^pQ^q\chi_{X^{p,q}}^{}$. Naively one would expect to identify the two superderivatives as a representation of the supercharges  -- the additional fermionic generators in superconformal algebra -- just as one treats the usual partial derivative $\partial_\mu$ as a representation of the momentum generator of the Poincare algebra. However, this identification fails: although the superderivatives do satisfy the same anticommutation relations, they are not simply the representation of supercharges. Recall the definition of superderivatives \eqref{Ds} and \eqref{anti relation}, from which we finds that superderivatives contain the usual partial derivative pieces. Formally one can write $D\sim Q+P, \overline{D}\sim \overline{Q}+P$, where $Q, \overline{Q}$ are supercharges and $P$ is the momentum generator. This realization indicates that $D$ and $\overline{D}$ are interwined with the usual partial derivatives, which makes the superconformal approach much more complicated, and so far we have not succeeded in using it as a starting point.

Nonetheless, there is something we can learn from superconformal algebra, encoded in five independent vanishing (anti)commutators:
\begin{align}
&\{\partial_\alpha,\partial_\beta\}=\{\partial_{\dot{\alpha}},\partial_{\dot{\beta}}\}=0,\nonumber \\
&[\partial_\alpha,\{\partial_\beta,\partial_{\dot{\alpha}}\}]=[\partial_{\dot{\alpha}},\{\partial_\alpha,\partial_{\dot{\beta}}\}]=0,\\
&[\{\partial_\alpha,\partial_{\dot{\alpha}}\},\{\partial_\beta,\partial_{\dot{\beta}}\}]=0. \nonumber 
\end{align}
The existence of the first branch makes use of the first identity, namely two consecutive superderivatives acting on symmetric indices automatically vanishes. To prove the existence of the second branch, we use the fact that superderivatives commutes with usual partial derivative, leading to the proof shown in \eqref{second branch proof}. The last identity is essential to prove the maps $l_n$ indeed satisfy the requirement when we consider the third branch. To summarize, the vanishing of these special combinations of superderivatives leads to the existence of three correction branches, just as the fact that one branch exists in non-supersymmetric theories can be connected to the single vanishing commutator $[\partial_\mu,\partial_\nu]=0$.  Future studies on the relation between the method we develop here and the superconformal algebra is very promising and interesting.

We have only analyzed a $N=1$ supersymmetry chiral theory so the technique we developed here should be easily generalized to other supersymmetric theories. In a forthcoming publication we will apply what we learn to vector superfields, and postpone extended supersymmetric theories to later studies.

\acknowledgments

We would like to thank Xiaochuan Lu for some useful comments on the draft. This work is partially supported by the National Science Foundation under Grant Number PHY-2112540.

\appendix

\section{Characters and Haar measures}
\label{app:characters}

Here we list a few characters in group $U(1),\ SU(2)$ and $SU(3)$ that are used in this paper. And we also list the related Haar measure for each group. For further characters, please refer to \cite{Hanany:2008sb}.

The characters are given by:
\begin{subequations}
\begin{align}
&\chi_{U(1)}=e^Q,\\
&\chi_{SU(2) fund}=z+\frac{1}{z},\\
&\chi_{SU(3) fund}=z_1+\frac{z_2}{z_1}+\frac{1}{z_2},
\end{align}
\end{subequations}
where $Q,\ z,\ z_1,\ z_2$ are related group parameters. In addition, the character of representation $(\frac{n}{2},0)$ and $(0,\frac{n}{2})$ are given by:
\begin{subequations}\label{n character}
\begin{align}
&\chi_{(\frac{n}{2},0)}=\frac{\sin((n+1)\Omega_x)}{\sin\Omega_x},\\
&\chi_{(0,\frac{n}{2})}=\frac{\sin((n+1)\Omega_y)}{\sin\Omega_y},
\end{align}
\end{subequations}
where $\Omega_{x,y}$ are defined through $x,y\equiv 2\cos(\Omega_{x,y})$, and $x,y$ are the related $SU(2)_L,SU(2)_R$ group parameters.

The Haar measures are given by: 
\begin{subequations}
\begin{align}
&\int d\mu_{U(1)}=\frac{1}{2\pi i}\oint_{|z|=1} \frac{dz}{z},\\
&\int d\mu_{SU(2)}=\frac{1}{2\pi i}\oint_{|z|=1} \frac{dz}{z}(1-z^2),\\
&\int d\mu_{SU(3)}=\frac{1}{(2\pi i)^2}\oint_{|z_1|=1} \frac{dz_1}{z_1}\oint_{|z_2|=1} \frac{dz_2}{z_2}(1-z_1z_2)(1-\frac{z_1^2}{z_2})(1-\frac{z_2^2}{z_1}).
\end{align}
\end{subequations}

\section{Proofs}

\subsection{Higher order corrections}\label{proof section}

In this Appendix, we show the detailed derivation leads to the explict form of coefficients in $l_n$ and $\overline{l}_n$ given in \eqref{eq:aandb}. Recall the definition of $l_n$ acting on a definite representation, which is given by:
\begin{equation}\label{ln}
(l_n)_{\dot{\alpha}}^{(\alpha_1\alpha_2\cdot\cdot\cdot\alpha_n)(\tau\beta_1\beta_2\cdot\cdot\cdot\beta_n)}Z^{\dot{\alpha}}_{(\alpha_1\alpha_2\cdot\cdot\cdot\alpha_n)}=(a_n\partial_{\dot{\alpha}}\partial_X+b_n\partial_X\partial_{\dot{\alpha}})\epsilon^{(X\alpha_1\alpha_2\cdot\cdot\cdot\alpha_n)(\tau\beta_1\beta_2\cdot\cdot\cdot\beta_n)}Z^{\dot{\alpha}}_{(\alpha_1\alpha_2\cdot\cdot\cdot\alpha_n)},
\end{equation}
where $\epsilon^{(X\alpha_1\alpha_2\cdot\cdot\cdot\alpha_n)(\tau\beta_1\beta_2\cdot\cdot\cdot\beta_n)}$ is defined to be the fully symmetrization permutations of its indices. For example, $\epsilon^{(\alpha_1\alpha_2)(\beta_1\beta_2)}=\epsilon^{\alpha_1\beta_1}\epsilon^{\alpha_2\beta_2}+\epsilon^{\alpha_2\beta_1}\epsilon^{\alpha_1\beta_2}$.

These maps should satisfy the following equation such that the above correction spaces obey the definition, i.e. $\{X\}^{n,1}$ corrects $\{X\}^{n-1,1}$ and $\{X^{n-1,0}\}$ with respect to $\{X\}^{n-2,0}$ when $n\geq3$:
\begin{equation}
[\partial_{\beta_n} (l_n)_{\dot{\alpha}}^{(\alpha_1\alpha_2\cdot\cdot\cdot\alpha_{n-1})(\tau\beta_1\beta_2\cdot\cdot\cdot\beta_n)}+(l_{n-1})_{\dot{\alpha}}^{(\alpha_1\alpha_2\cdot\cdot\cdot\alpha_{n-1})(\tau\beta_1\beta_2\cdot\cdot\cdot\beta_{n-1})}\partial^{\alpha_n}]Z^{\dot{\alpha}}_{(\alpha_1\alpha_2\cdot\cdot\cdot\alpha_n)}=0
\end{equation}
Expand it using \eqref{ln} and we get
\begin{equation}
\begin{split}
&[(a_n\partial_{\beta_n}\partial_{\dot{\alpha}}\partial_X+b_n\partial_{\beta_n}\partial_X\partial_{\dot{\alpha}})\epsilon^{(X\alpha_1\alpha_2\cdot\cdot\cdot\alpha_n)(\tau\beta_1\beta_2\cdot\cdot\cdot\beta_n)}\\
&+(a_{n-1}\partial_{\dot{\alpha}}\partial_X\partial^{\alpha_n}+b_{n-1}\partial_X\partial_{\dot{\alpha}}\partial^{\alpha_n})\epsilon^{(X\alpha_1\alpha_2\cdot\cdot\cdot\alpha_{n-1})(\tau\beta_1\beta_2\cdot\cdot\cdot\beta_{n-1})}]Z^{\dot{\alpha}}_{(\alpha_1\alpha_2\cdot\cdot\cdot\alpha_n)}=0
\end{split}
\end{equation}
Use the identity $[\partial^{\alpha_n},\{\partial_{\dot{\alpha}},\partial_X\}]=0$, we can rewrite the second term as:
\begin{equation}
\begin{split}
&(a_{n-1}\partial_{\dot{\alpha}}\partial_X\partial^{\alpha_n}+b_{n-1}\partial_X\partial_{\dot{\alpha}}\partial^{\alpha_n})\epsilon^{(X\alpha_1\alpha_2\cdot\cdot\cdot\alpha_{n-1})(\tau\beta_1\beta_2\cdot\cdot\cdot\beta_{n-1})}\\
&=[a_{n-1}(\partial^{\alpha_n}\partial_{\dot{\alpha}}\partial_X+\partial^{\alpha_n}\partial_X\partial_{\dot{\alpha}})+(b_{n-1}-a_{n-1})\partial_X\partial_{\dot{\alpha}}\partial^{\alpha_n}]\\
&*\sum_i\epsilon^{X\beta_i}\epsilon^{(\alpha_1\alpha_2\cdot\cdot\cdot\alpha_{n-1})(\tau\beta_1\beta_2\cdot\cdot\cdot\beta_{n-1}/\beta_i)}\\
&=-\sum_i[a_{n-1}(\partial^{\alpha_n}\partial_{\dot{\alpha}}\partial^{\beta_i}+\partial^{\alpha_n}\partial^{\beta_i}\partial_{\dot{\alpha}})+(b_{n-1}-a_{n-1})\partial_X\partial_{\dot{\alpha}}\partial^{\alpha_n}]\epsilon^{(\alpha_1\alpha_2\cdot\cdot\cdot\alpha_{n-1})(\tau\beta_1\beta_2\cdot\cdot\cdot\beta_{n-1}/\beta_i)},
\end{split}
\end{equation}
where $(\tau\beta_1\beta_2\cdot\cdot\cdot\beta_{n-1}/\beta_i)$ denotes that the indices do not contain the $\beta_i$ piece.
We can separate it into two parts and compare their coefficients and solve for $a_n$ and $b_n$. The two equations are given by
\begin{equation}\label{first equation}
\begin{split}
&a_n\partial_{\beta_n}\partial_{\dot{\alpha}}\partial_X\epsilon^{(X\alpha_1\alpha_2\cdot\cdot\cdot\alpha_n)(\tau\beta_1\beta_2\cdot\cdot\cdot\beta_n)}\\
&=\sum_i[a_{n-1}\partial^{\alpha_n}\partial_{\dot{\alpha}}\partial^{\beta_i}+(b_{n-1}-a_{n-1})\partial^{\beta_i}\partial_{\dot{\alpha}}\partial^{\alpha_n}]\epsilon^{(\alpha_1\alpha_2\cdot\cdot\cdot\alpha_{n-1})(\tau\beta_1\beta_2\cdot\cdot\cdot\beta_{n-1}/\beta_i)}
\end{split}
\end{equation}
and
\begin{equation}\label{second equation}
b_n\partial_{\beta_n}\partial_X\partial_{\dot{\alpha}}\epsilon^{(X\alpha_1\alpha_2\cdot\cdot\cdot\alpha_n)(\tau\beta_1\beta_2\cdot\cdot\cdot\beta_n)}=\sum_ia_{n-1}\partial^{\alpha_n}\partial^{\beta_i}\partial_{\dot{\alpha}}\epsilon^{(\alpha_1\alpha_2\cdot\cdot\cdot\alpha_{n-1})(\tau\beta_1\beta_2\cdot\cdot\cdot\beta_{n-1}/\beta_i)}
\end{equation}
We can expand the $\epsilon$ symbol using the identity $\epsilon_{AB}\epsilon_{CD}+\epsilon_{AC}\epsilon_{DB}+\epsilon_{AD}\epsilon_{BC}=0$:
\begin{equation}
\begin{split}
&\epsilon^{(X\alpha_1\alpha_2\cdot\cdot\cdot\alpha_n)(\tau\beta_1\beta_2\cdot\cdot\cdot\beta_n)}\\&=\sum_i(\epsilon^{X\beta_n}\epsilon^{\alpha_n\beta_i}+n\epsilon^{X\beta_i}\epsilon^{\alpha_n\beta_n})\epsilon^{(\alpha_1\alpha_2\cdot\cdot\cdot\alpha_{n-1})(\tau\beta_1\beta_2\cdot\cdot\cdot\beta_{n-1}/\beta_i)}\\
&=\sum_i(\epsilon^{X\alpha_n}\epsilon^{\beta_n\beta_i}+(n+1)\epsilon^{X\beta_i}\epsilon^{\alpha_n\beta_n})\epsilon^{(\alpha_1\alpha_2\cdot\cdot\cdot\alpha_{n-1})(\tau\beta_1\beta_2\cdot\cdot\cdot\beta_{n-1}/\beta_i)}
\end{split}
\end{equation}
Put this back into the first equation, we get:
\begin{equation}
\begin{split}
&a_n\partial_{\beta_n}\partial_{\dot{\alpha}}\partial_X\epsilon^{(X\alpha_1\alpha_2\cdot\cdot\cdot\alpha_n)(\tau\beta_1\beta_2\cdot\cdot\cdot\beta_n)}\\
&=\sum_i[a_n\partial^{\beta_i}\partial_{\dot{\alpha}}\partial^{\alpha_n}-(n+1)a_n\partial^{\alpha_n}\partial_{\dot{\alpha}}\partial^{\beta_i}]\epsilon^{(\alpha_1\alpha_2\cdot\cdot\cdot\alpha_{n-1})(\tau\beta_1\beta_2\cdot\cdot\cdot\beta_{n-1}/\beta_i)}\\
&=\sum_i[a_{n-1}\partial^{\alpha_n}\partial_{\dot{\alpha}}\partial^{\beta_i}+(b_{n-1}-a_{n-1})\partial^{\beta_i}\partial_{\dot{\alpha}}\partial^{\alpha_n}]\epsilon^{(\alpha_1\alpha_2\cdot\cdot\cdot\alpha_{n-1})(\tau\beta_1\beta_2\cdot\cdot\cdot\beta_{n-1}/\beta_i)}
\end{split}
\end{equation}
from which we can solve for $a_n$ and $b_n$ recursively:
\begin{equation}
a_n=b_{n-1}-a_{n-1},\ \ -(n+1)a_n=a_{n-1}
\end{equation}
The initial condition is given by $a_0=2$ and $b_0=1$\footnote{In principle the choice is not unique and one can always multiply this solution by a common factor.} in order to generate the same $l_0$ define in Section \ref{second branch}. . Then we can solve $a_n$ and $b_n$ for general n:
\begin{equation}
a_n=(-1)^n\frac{2}{(n+1)!},\ \ b_n=(-1)^n\frac{2}{(n+2)n!}.
\end{equation}
We only make use of the first equation of \eqref{first equation} and \eqref{second equation}. Now we need to prove that this solution does satisfy the second one \eqref{second equation}:
\begin{equation}
\begin{split}
LHS&=(-1)^n\frac{2}{(n+2)n!}\partial_{\beta_n}\partial_X\partial_{\dot{\alpha}}\epsilon^{(X\alpha_1\alpha_2\cdot\cdot\cdot\alpha_n)(\tau\beta_1\beta_2\cdot\cdot\cdot\beta_n)}\\
&=(-1)^n\frac{2}{(n+2)n!}\partial_{\beta_n}\partial_X\partial_{\dot{\alpha}}\sum_i(\epsilon^{X\beta_n}\epsilon^{\alpha_n\beta_i}+n\epsilon^{X\beta_i}\epsilon^{\alpha_n\beta_n})\epsilon^{(\alpha_1\alpha_2\cdot\cdot\cdot\alpha_{n-1})(\tau\beta_1\beta_2\cdot\cdot\cdot\beta_{n-1}/\beta_i)}\\
&=\sum_i(-1)^{n+1}\frac{2}{(n+2)n!}(\partial_X\partial^X\partial_{\dot{\alpha}}\epsilon^{\alpha_n\beta_i}+n\partial^{\alpha_n}\partial^{\beta_i}\partial_{\dot{\alpha}})\epsilon^{(\alpha_1\alpha_2\cdot\cdot\cdot\alpha_{n-1})(\tau\beta_1\beta_2\cdot\cdot\cdot\beta_{n-1}/\beta_i)}\\
RHS&=\sum_i(-1)^{n-1}\frac{2}{(n)!}\partial^{\alpha_n}\partial^{\beta_i}\partial_{\dot{\alpha}}\epsilon^{(\alpha_1\alpha_2\cdot\cdot\cdot\alpha_{n-1})(\tau\beta_1\beta_2\cdot\cdot\cdot\beta_{n-1}/\beta_i)}
\end{split}
\end{equation}
$LHS=RHS$ follows from the fact that $\partial_X\partial^X\epsilon^{\alpha_n\beta_i}=2\partial^{\alpha_n}\partial^{\beta_i}$.

Futhermore, we require that $l_{n+1}l_n=0$\footnote{This allows the horizontal composite maps to vanish by virtue of the definition of correction spaces.}, and the proof is straight forward.
\begin{equation}
\begin{split}
&(l_{n+1})_{\dot{\alpha}(\beta_1\beta_2\cdot\cdot\cdot\beta_{n+1})(\gamma_1\gamma_2\cdot\cdot\cdot\gamma_{n+2})}(l_n)^{\dot{\alpha}(\alpha_1\alpha_2\cdot\cdot\cdot\alpha_n)(\beta_1\beta_2\cdot\cdot\cdot\beta_{n+1})}\\
&\propto(\frac{1}{n+2}\partial_{\dot{\alpha}}\partial^X+\frac{1}{n+3}\partial^X\partial_{\dot{\alpha}})(\frac{1}{n+1}\partial^{\dot{\alpha}}\partial_Y+\frac{1}{n+2}\partial_Y\partial^{\dot{\alpha}})\\
&*\epsilon_{(X\beta_1\beta_2\cdot\cdot\cdot\beta_{n+1})(\gamma_1\gamma_2\cdot\cdot\cdot\gamma_{n+2})}\epsilon^{(Y\alpha_1\alpha_2\cdot\cdot\cdot\alpha_n)(\beta_1\beta_2\cdot\cdot\cdot\beta_{n+1})}\\
&=(\frac{1}{n+2}\partial_{\dot{\alpha}}\partial^X+\frac{1}{n+3}\partial^X\partial_{\dot{\alpha}})(\frac{1}{n+1}\partial^{\dot{\alpha}}\partial_Y+\frac{1}{n+2}\partial_Y\partial^{\dot{\alpha}})\\
&*\sum_{i\neq j}(n+2)\epsilon_{X\gamma_i}\delta^Y_{\gamma_j}\epsilon_{(\beta_1\beta_2\cdot\cdot\cdot\beta_{n+1}/\beta_j)(\gamma_1\gamma_2\cdot\cdot\cdot\gamma_{n+2}/\gamma_j)}\epsilon^{(\alpha_1\alpha_2\cdot\cdot\cdot\alpha_n)(\beta_1\beta_2\cdot\cdot\cdot\beta_{n+1}/\beta_j)}
\end{split}
\end{equation}
Therefore to prove the above expression is 0, it suffices to prove the following equation
\begin{equation}
(\frac{1}{n+2}\partial_{\dot{\alpha}}\partial_X+\frac{1}{n+3}\partial_X\partial_{\dot{\alpha}})(\frac{1}{n+1}\partial^{\dot{\alpha}}\partial_Y+\frac{1}{n+2}\partial_Y\partial^{\dot{\alpha}})+(X\leftrightarrow Y)=0
\end{equation}
Expand the LHS and terms involving $\partial_X\partial_Y$ vanishes due to its antisymmetric property. In addition, we use the defining anticommutator to rewrite $\partial_{\dot{\alpha}}\partial_X$ as $\partial_{\dot{\alpha}X}-\partial_X\partial_{\dot{\alpha}}$. Finally we are left with
\begin{equation}
\begin{split}
&(\frac{1}{n+2}\partial_{\dot{\alpha}}\partial_X+\frac{1}{n+3}\partial_X\partial_{\dot{\alpha}})(\frac{1}{n+1}\partial^{\dot{\alpha}}\partial_Y+\frac{1}{n+2}\partial_Y\partial^{\dot{\alpha}})+(X\leftrightarrow Y)\\
&=\frac{1}{(n+1)(n+2)}[\partial_{\alpha\dot{\alpha}}\{\partial^\alpha,\partial_{\dot{\beta}}\}+\partial_{\alpha\dot{\beta}}\{\partial^\alpha,\partial_{\dot{\alpha}}\}]\\
&=\frac{1}{(n+1)(n+2)}[\{\partial^\alpha,\partial_{\dot{\beta}}\},\{\partial_\alpha,\partial_{\dot{\alpha}}\}]=0,
\end{split}
\end{equation}
which identically vanishes. This completes the proof of $l_{n+1}l_n=0$.

\subsection{Starting space of second branch}\label{starting space}
Here we answer the question arised at the end of \ref{second branch}, where one may wonder why the second branch starts from $\{X^{\dot{\alpha}}\}^{2,1}$. The general solution to this problem is to consider the nearest spaces and check if they can satisfy the definition. If not, we need to move to the second nearest space and check again. Specifically, we know that $\{X^{\alpha}\}^{1,0}$ gives the IBP relations, and the nearest spaces are $\{X\}^{2,0}$ and $\{X^{\alpha\dot{\alpha}}\}^{1,1}$. The first one already exists in the first branch, while the second one cannot satisfy the definition.  We can prove this by looking at the equation
\begin{equation}
(a\partial_\alpha\partial_{\dot{\alpha}}+b\partial_{\dot{\alpha}}\partial_\alpha)X^{\alpha\dot{\alpha}}=0.
\end{equation}
If the space $\{X^{\alpha\dot{\alpha}}\}^{1,1}$ is indeed a correction, then we must find a nontrivial solution $(a,b)$ of the above equation. However this is impossible and therefore leads to contradiction. Then we do the same for the second nearest spaces $\{X^{\dot{\alpha}}\}^{2,1}$ and we find that when it transforms under $(0,\frac{1}{2})$, it satisfies the definition. The reason it doesn't transform under $(1,\frac{1}{2})$ is because in that case it leads to null space under the map $\partial_{\dot{\alpha}}\partial_{\dot{\beta}}X^{(\dot{\alpha}\dot{\beta})\alpha}=0$ and therefore cannot be a correction. Now we have completed the proof why the second branch starts from $\{X^{\dot{\alpha}}\}^{2,1}$.

\subsection{Infinite sum}
The infinite sum is given by:
\begin{equation}\label{formalsum1}
\begin{split}
\sum P^pQ^q\chi_{X^{p,q}}^{}&=1-P^2Q^2\\
&+\sum_{n=1}((-P)^n\Lambda_n+(-Q)^n\overline{\Lambda}_n)\\
&+PQ\sum_{n=0}(Py(-P)^n\Lambda_n+Qx(-Q)^n\overline{\Lambda}_n)\\
&+P^2Q^2\sum_{n=0}(P^2(-P)^n\Lambda_n+Q^2(-Q)^n\overline{\Lambda}_n)\\
&=-1-P^2Q^2+\sum_{n=0}((-P)^n\Lambda_n+(-Q)^n\overline{\Lambda}_n)(1+P^2Qy+PQ^2x+P^4Q^2+P^2Q^4)
\end{split}
\end{equation}
where $\Lambda_n$ and $\overline{\Lambda}_n$ are the characters of $(\frac{n}{2},0)$ and $(0,\frac{n}{2})$ given in \eqref{n character}.

\bibliographystyle{utphys}
\bibliography{SUSY-HS}

\end{document}